\documentclass[preprint2]{aastex} 

\usepackage[latin1]{inputenc} 
\usepackage{graphicx} 

\slugcomment{To appear in PASP} 
\shorttitle{The 2DPHOT environment}
\shortauthors{F. La Barbera et al.}

\author{F. La Barbera}
\affil{INAF-Osservatorio Astronomico di Capodimonte, via Moiariello
  n.16, Napoli, Italy}
\email{labarber@na.astro.it}
\author{R. R. de Carvalho}
\affil{INAF-VSTCeN, via Moiariello
  n.16, Napoli, Italy\footnote{On leave of absence INPE/DAS, Av. dos Astronautas 1758, S$\tilde{a}$o Jos\'e dos Campos, SP 12227-010, Brazil.}}
\email{reinaldo@das.inpe.br}
\author{J.L. Kohl-Moreira}
\affil{Observat\'orio Nacional, Rua General Jos\'e Cristino 77, S$\tilde{a}$o Crist\'ov$\tilde{a}$o, Rio de Janeiro 20921-400, Brazil}
\author{R.R. Gal}
\affil{University of Hawaii, Institute for Astronomy, 2680 Woodlawn Dr., Honolulu, HI, 96822, United States}
\author{M. Soares-Santos}
\affil{Instituto de Astronomia, Geof\'isica e Ciencias Atmosf\'ericas}
\author{M. Capaccioli}
\affil{INAF-VSTCeN, via Moiariello
  n.16, Napoli, Italy}
\author{R. Santos}
\affil{INPE/LAC, Av. dos Astronautas 1758, S$\tilde{a}$o Jos\'e dos Campos, SP 12227-010, Brazil}
\author{N. Sant'Anna}
\affil{INPE/LAC, Av. dos Astronautas 1758, S$\tilde{a}$o Jos\'e dos Campos, SP 12227-010, Brazil}

\begin{document}

\title{2DPHOT:  a multi-purpose  environment  for  the
  two-dimensional analysis of wide-field images}


%
%
%
%
%
%

\begin{abstract}
  We  describe  2DPHOT, a  general  purpose  analysis environment  for
  source  detection and analysis in deep wide-field images.
  2DPHOT is  an automated tool  to obtain both integrated  and surface
  photometry  of   galaxies  in  an  image,  to   perform  reliable
  star-galaxy separation  with accurate estimates  of contamination at
  faint flux levels, and to  estimate completeness of the image catalog.
  We  describe the  analysis strategy  on which
  2DPHOT is based, and provide a detailed description of the different
  algorithms  implemented in  the  package.  This  new environment  is
  intended  as a dedicated  tool to  process the  wealth of  data from
  wide-field imaging surveys. To this end, the package is complemented
  by  2DGUI, an environment that  allows multiple
  processing of data using a range of computing architectures.
\end{abstract}


\section{Introduction}\label{intro}

In the past decade, wide-field surveys have provided the scientific
community with a huge amount of spectroscopic and photometric data,
allowing significant progress in our understanding of the Universe.
Perhaps the most widely known example is the Sloan Digital Sky Survey
(SDSS), whose sixth data release has now provided photometry in five
bands for more than $2 \! \cdot \!  10^8$ astronomical objects, as
well as spectra of about one million sources (see
\citealt{Adelman:07}) over more than $8500$ square degrees on the sky.
One key to the success of the SDSS has been its capability to
effectively store, process, and analyze, in a fully automated fashion,
the vast amount of data gathered during survey operations.  This goal
was achieved by using dedicated and well-designed software pipelines,
updated during survey operations with
reprocessing for the delivery of new data releases. In the coming
years, many general purpose astronomical surveys are slated to begin
taking data.  These wide-field imaging projects will gather deeper and
deeper multi-waveband data over large sky areas, producing ever
greater data flows. The scientific community must manage and analyze
the huge wealth of information contained in these enormous datasets.

\begin{figure*}
\begin{center}
\scalebox{0.53}{\includegraphics{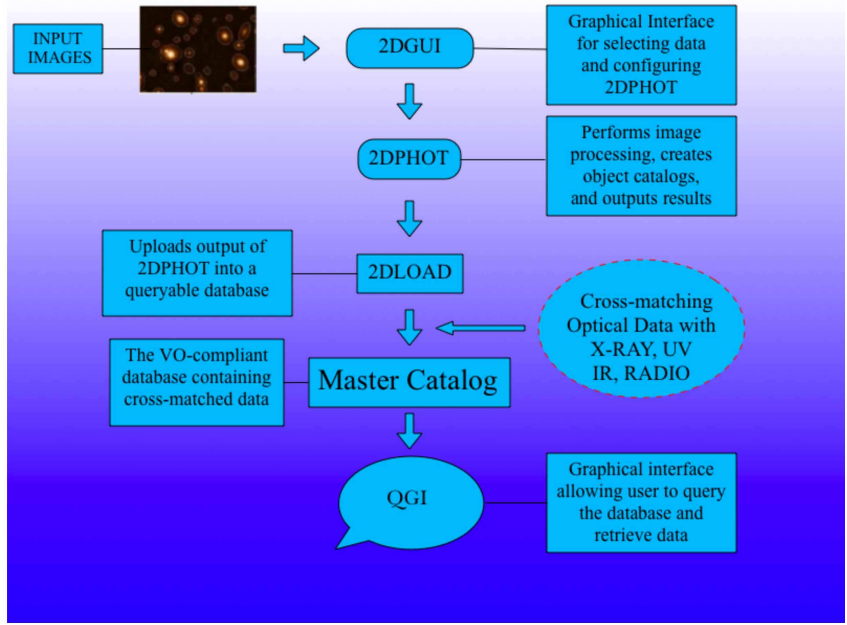}}
\end{center}
\caption[]{\footnotesize   Schematic representation of the 2DPHOT
  environment.~\label{DBFIG} }
\end{figure*}

In this environment, we have undertaken the development of a new image
analysis tool called 2DPHOT, designed to derive two-dimensional
information by analyzing both the surface brightness distributions of
individual astronomical sources and the spatial distribution of these
sources in the image.  The package includes several tasks, such as
star/galaxy classification, measurement of both integrated and surface
photometry of galaxies, PSF modeling, and estimation of catalog
completeness and classification accuracy. The package is complemented by a
graphical interface named 2DGUI. A schematic view of the 2DPHOT environment is shown in
Fig.~\ref{DBFIG}.  Briefly, the environment is conceived as
follows. To start processing, the input images are uploaded to a
computer system (e.g.  a local cluster or a grid computer) via the
2DGUI interface.  2DGUI also allows the user to configure the 2DPHOT
input parameters. A scheduler is also included, allowing timed and
sequential execution of several 2DPHOT runs to be performed on the
same computer.  The actual image analysis is done by the 2DPHOT
package, which is the core of the whole environment.  During
execution, several output tables and plots are produced, showing the
different steps of the image analysis and providing a means of quality
control. These data, which can also be directly downloaded from the
user through the 2DGUI interface, are all uploaded into a database
system (by the 2DLOAD application, see Fig.~\ref{DBFIG}).  This system
produces a master catalog, by cross-matching 2DPHOT output results
with information provided from other VO-compliant web services, and
allows the user to perform data queries on this master catalog.

There are several survey projects for which the 2DPHOT environment has
already  been partly  implemented  or will  be  implemented. We  have
automatically  processed the  $g$- and  $r$-band images  from  the Palomar
Abell Cluster  Survey~\citep{Gal:00}, with the main  goal of measuring
structural parameters,  i.e.  the effective  radius, the corresponding
mean  surface brightness,  and the  S\'{e}rsic  index $n$  of galaxies  in
clusters with different  richnesses, in the redshift range  of 0.05 to
0.2.  The  structural  parameters  have  been  used  to  estimate  the
environmental dependencies  of internal color  gradients in early-type
galaxies  (see \citealt{LaB:05}).   Some examples  of  general purpose
imaging surveys  to be analyzed by  2DPHOT are those  carried out with
the  VLT Survey  Telescope (VST),  a 2.6m  diameter  imaging telescope
equipped with a large format (16k x16k pixels) CCD camera yielding a 1
square degree  field of view.  The  VST, which will be  located at the
ESO  Cerro   Paranal  Observatory  (Chile),  has   been  designed  and
constructed  under  a  joint   venture  of  ESO  and  the  Capodimonte
Astronomical  Observatory  (OAC).   Several  survey projects  will  be
carried      out      with      Capodimonte's      VST      guaranteed
time\footnote{\footnotesize See http://vstportal.oacn.inaf.it/}.

One  of the  most interesting  science  cases for  the development  of
2DPHOT    is    the     Kilo-Degree    Survey    with    VST    (KIDS,
see~\citealt{Arn:07}), a  public survey project which  will image 1500
square  degrees of  the southern  sky in  the $u$$g$$r$$i$  bands.  As
shown  in a  forthcoming paper  \citep{paper2:08},
applying 2DPHOT  to a moderately  deep survey such as  VST-KIDS allows
detection and measurement of massive galaxy clusters up to redshift $z
\sim 1.2$ with high  completeness.  This cluster abundance measurement
can be used  to set strong constraints on the  dark energy equation of
state, which is one of the most crucial issues of modern observational
and  theoretical  cosmology.   Reliable star-galaxy  separation,  with
accurate estimates of contamination at very faint flux levels, as well
as  an  accurate cluster  detection  algorithm  are  among the  2DPHOT
features of paramount importance for such a dark energy project.

This  paper  presents  the  2DPHOT  package\footnote{\footnotesize The  source code
  of the package  is available on request to the authors in its
  standard form, namely without the VO structure and 2DGUI interface,
which will be made available in the near future. },
describing the image  analysis strategy on which it  is based, as well
as all  the algorithms which  are implemented for the  different tasks
the  package  performs.   We   also  describe  briefly  the  web-based
graphical interface.  This  paper is intended as a  reference work for
all current  and forthcoming  scientific applications of  2DPHOT.  The
layout  of the paper  closely follows  the order  of execution  of the
2DPHOT   tasks.   In   Sec.~\ref{2DPHOT},  we   give   general,  short
descriptions  of these  tasks, and  how they  are linked  during image
analysis.  Section~\ref{catalog}  describes the initial components of
the first analysis  step, i.e.  how 2DPHOT produces  the image catalog
and identifies those objects which are classified as sure stars in the
input image.  The analysis of  each source in the catalog is performed
by extracting a  stamp image from the input  frame and constructing a
corresponding mask file (Sec.~\ref{stamps}).  The package performs PSF
modeling and  derives rough structural  parameters for all  sources in
the   image   as   described  in   Secs.~\ref{PSF}   and~\ref{INI2DF},
respectively.    Sec.~\ref{SGCLAS}    deals   with   the   star/galaxy
separation,  while  Sec.~\ref{2DFIT} describes  the  final fitting  of
galaxy stamps with  seeing-convolved S\'{e}rsic models.  The isophotal
analysis of galaxy stamps is then described in Sec.~\ref{SPHOT}, while
the determination  of the seeing-corrected  galaxy aperture magnitudes
is  outlined in  Sec.~\ref{GROWTH}.  Sections  ~\ref{COMPLETENESS} and
~\ref{SG_CONTAM} describe how 2DPHOT estimates the completeness of the
galaxy  catalog and  the  uncertainty in  the star/galaxy  separation.
Section~13 shows how 2DPHOT performs in estimating contamination and 
completeness at faint magnitudes.
 Finally, Sec.~\ref{INTERF}  presents the
graphical  interface  (2DGUI).    A  summary  is  given  in
Sec.~\ref{SUMMARY}.   The  input   parameters  and  output  quantities
measured   by   2DPHOT    are   provided   in   Appendices~\ref{INPAR}
and~\ref{OUTPAR}, respectively.

\section{The 2DPHOT package: tasks and analysis strategy}
\label{2DPHOT}
2DPHOT is  designed to have a  simple structure consisting  of a shell
script running  a suite of  C and Fortran77 programs developed using
freely available software libraries.  2DPHOT works on both single-chip
and wide-field (up to 16000x16000 pixels) images with a set of
input parameters provided either at the invocation of
the shell script through a command line syntax or a corresponding
graphical interface (see Sec.~\ref{INTERF}).  Thus, the package can be
used either as a standalone  application or via a dedicated web-based
interface.   The  list of  input  parameters along with short
descriptions is provided in Appendix~\ref{INPAR}.

The main tasks of 2DPHOT are:
\begin{description}
\item[1.]  Producing a cleaned catalog of the image.
\item[2.]  Performing reliable star/galaxy  classification.
\item[3.]  Estimating  the completeness of the galaxy  catalog and the
  contamination due to star/galaxy misclassification.
\item[4.] Constructing an accurate  model of the Point Spread Function
  (PSF)  of the  input  image, taking  into  account possible  spatial
  variations of  the PSF  as well as  deviations of  stellar isophotes
  from circularity.
\item[5.]  Deriving  structural parameters  of  galaxies by  fitting
  galaxy  images  with  two-dimensional  PSF-convolved  S\'{e}rsic  models.
\item[6.]  Measuring galaxy isophotes by fitting them with
  Fourier-expanded ellipses, and derivation of one-dimensional surface
  brightness profiles of galaxies.
\item[7.]  Measuring the  growth  curve of  seeing corrected  aperture
  magnitudes of galaxies.
\end{description}

\begin{figure*}
\begin{center}
\scalebox{1.0}{\includegraphics{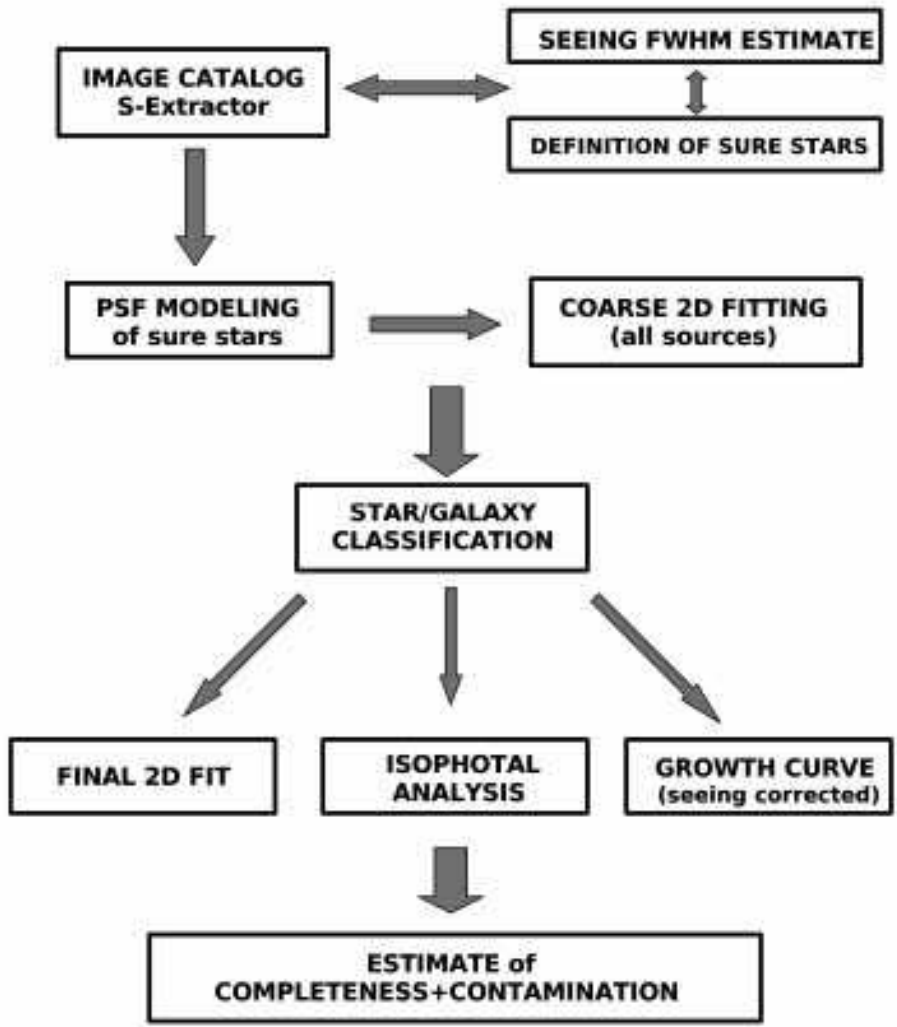}}
\end{center}
\caption[]{\footnotesize   Image   analysis   flow   of   2DPHOT.~\label{FLOW} }
\end{figure*}

All  of these  tasks  are part  of  an image  analysis  flow and  are
strictly linked with each other such that the output from one
task  is used  as input  to the  subsequent tasks.  Figure~\ref{FLOW}
provides  a graphical  representation of  this flow,  where  the boxes
represent different steps in the image analysis and the arrows follow
the  image  processing  timeline.    The  package  starts  by  running
S-Extractor~\citep{BeA:96}  on the  input image  through  an iterative
procedure, allowing  simultaneous measurement  of the seeing  FWHM and
removal of spurious object detections.  Stamp and mask images are then
extracted  for each object  in the  cleaned catalog,  and are  used to
model the PSF across the field  and to obtain a coarse estimate of the
S\'{e}rsic parameters of the detected sources.  Using both the S-Extractor
stellarity  index and  the coarse  effective radius  estimates, 2DPHOT
performs star/galaxy classification.  The selected  galaxies are
then analyzed using  a two-dimensional fitting procedure as  well as a
full isophotal analysis. Seeing corrected aperture magnitudes are also
estimated.   At  this point,  simulations  are  performed to  estimate
completeness   and  contamination   of  the   final catalog. 

The  following sections  describe  all of  the  image analysis  steps,
following  the  diagram  in  Fig.~\ref{FLOW}.  The  output  quantities
measured by  2DPHOT are  summarized in Appendix~\ref{OUTPAR}.

\section{The image catalog and the definition  of `sure stars'}
\label{catalog}
\begin{figure}
\begin{center}
\scalebox{0.41}{\includegraphics{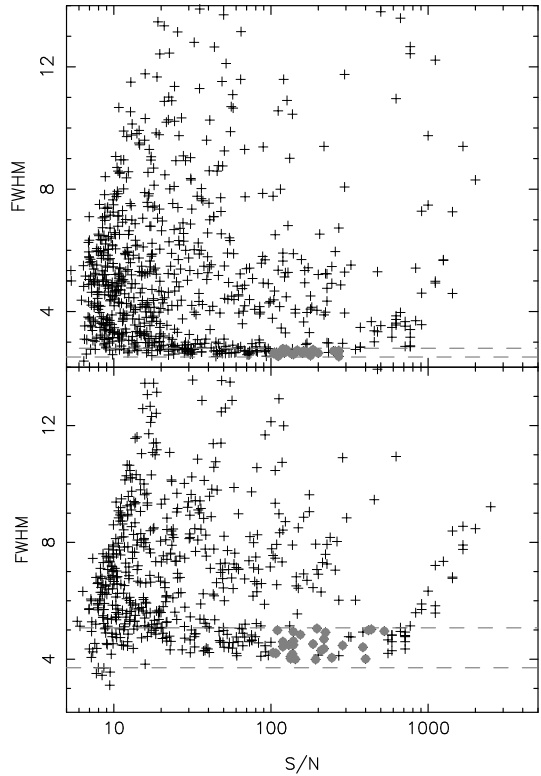}}
\end{center}
\caption[]{\footnotesize Locus of sure  stars for two $g$-band images from
  the  Palomar Abell Cluster  Survey (see  Gal et  al.  2000).   The  images  were  obtained with  a  SITe  2048$\times$2048,
  AR-coated CCD, at  the Palomar 1.5m telescope, and  cover an area of
  12.56$^{\prime}\times$12.56$^{\prime}$   around   the   cluster   of
  galaxies Abell  2495 (at $z\sim0.09$),  with a pixel scale  of 0.368
  arcsec/pixel. Both panels plot the  $FWHM$ vs. $S/N$ diagram for all
  the sources  in the same field  around the cluster  center, with the
  lower  panel  for  the   image  taken  in  worse  seeing  conditions
  ($FWHM\sim1.7''$).  The  $FWHM$ is given in pixel  units.  The $S/N$
  ratio  is  computed  as  the  inverse  of  the  uncertainty  on  the
  S-Extractor Kron magnitude (neglecting readout noise).  The locus of
  sure stars is  defined by the two horizontal  dashed gray lines that
  mark  the $\pm2\sigma$  region around  the  peak value  of the  FWHM
  distribution  of star  candidates.  Sure  stars are  defined  as the
  bright  ($S/N>100$) non-saturated star  candidates which  lie inside
  the  sure  star  locus,  and   are  plotted  as  grey  circles  (see
  text). \label{surestars} }
\end{figure}

2DPHOT produces the source catalog from the input image using the S-Extractor
package~\citep{BeA:96}.  Star/galaxy separation is performed on the
basis of the S-Extractor stellarity index $SI$ and the effective
radius parameter $r_e$ (see Sec.~\ref{whyparam}).  In order to obtain a
reliable estimate of the stellarity index, the seeing FWHM of the
input image has to be provided to S-Extractor via the $SEEING\_FWHM$
input parameter (see the S-Extractor
documentation\footnote{\footnotesize $http://terapix.iap.fr/rubrique.php?id\_rubrique=91/index.html$}).
To measure this, 2DPHOT produces a catalog from the input image via the
following two-step procedure.  S-Extractor is first run for the sole
purpose of detecting sources in the input image and calculating their
Kron magnitudes and FWHM and ELLIPTICITY parameters.  By applying a
$3\sigma$ clipping procedure to the FWHM and ELLIPTICITY distributions
of all the bright ($S/N > 100$) unsaturated objects, 2DPHOT generates
a preliminary list of candidate stars.  The peak value $f$ and the
width $\sigma$ of the FWHM distribution of these objects is derived
using the bi-weight estimator (Beers et al.~1990).  The values of $f$
and $\sigma$ define what we call the {\it sure star locus}, with the
{\it sure stars} being the objects that lie within $\pm 2 \sigma$ of
$f$.  Given the values of $f$ and $\sigma$, S-Extractor is run a
second time by setting the $SEEING\_FWHM$ parameter to the value of
$f$.  As an example of this procedure, Fig.~\ref{surestars} shows the
FWHM versus $S/N$ ratio diagram for all the detected sources in two
CCD images of the galaxy cluster Abell 2495, which has been observed
twice, under different seeing conditions, as part of the Palomar Abell
Cluster Survey (Gal et al.  2000). Similar figures showing the sure
star locus and the sure stars are automatically produced during each
run of 2DPHOT.

After  sure stars  are defined,  the  catalog is  cleaned of  spurious
detections  by excluding  all sources  3$\sigma$ below  the  sure star
locus.  Objects whose distance from the image edges, in units of their
FWHM value, is  smaller than {\it REDGE}, where {\it  REDGE} is one of
the input  2DPHOT parameters (see Appendix~A), are  also excluded from
the  analysis   since  their  photometry  can   be  incomplete  and/or
corrupted.

\section{Extraction of stamps and  mask images}
\label{stamps}
\begin{figure*}
\begin{center}
\scalebox{0.5}{\includegraphics{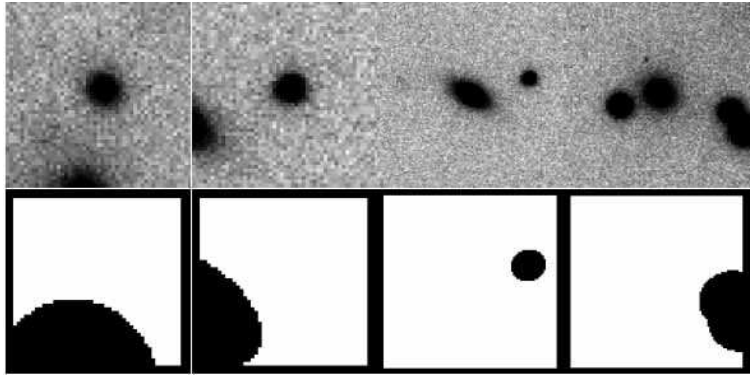}}
\end{center}
\caption[]{\footnotesize Examples of stamp and mask images produced by
2DPHOT.  Stamps are shown in the upper panels, while lower panels plot
the corresponding  mask images.  The  pixels which are flagged  in the
mask files are plotted in black.  Notice that pixels very close to the
stamp edges  are also  flagged in  each mask image.   This is  done to
reduce     computational     overhead     in    the     2D     fitting
algorithm. \label{masks} }
\end{figure*}

For each  detected source, 2DPHOT extracts an image  section (stamp) centered
on the  source.  The area of  the stamp is
proportional to the  $ISOAREA$ output by S-Extractor such  that a wide
sky region  around the central object  is also included  in the stamp.
This allows a reliable estimate of the local background to be obtained
from the two-dimensional  fitting program (see Sec.~\ref{2DFIT}).  For
each stamp, a  mask image is also produced by  flagging all the pixels
that  belong  to  all the  other  sources  in  the input  image  whose
isophotal areas overlap  the given stamp.  The  isophotal areas are
defined  through  the  $ISOAREA$,  $ELLIPTICITY$, and  position  angle
($PA$) parameters  from S-Extractor, by multiplying  the $ISOAREA$ value
by an expansion factor $EXPND$ (with a default value of 1.5), which is
an input parameter of 2DPHOT.   This expansion factor allows us to mask
also  the  faintest  diffuse  external regions  of  each  object.
Sources whose isophotal areas overlap  the central source by more than
$50\%$ are  not masked  out and are  analyzed simultaneously  with the
central object (see Secs.~\ref{INI2DF} and~\ref{2DFIT}).  For each stamp,
the  number of  sources  treated simultaneously  is  written into  the
$NOBJ$  keyword of  the  corresponding mask  file  header.  The  local
background  value and  its standard  deviation are  also  estimated by
applying biweight statistics to all  the pixels which do not belong to
the isophotal  area of the central  source and are not  flagged in the
mask file. These values are stored in the keywords $M\_BK$ and $S\_BK$
of the  mask file  header, respectively.  Fig.~\ref{masks}  shows some
examples  of  the stamp  and  mask  images  automatically produced  by
2DPHOT.

\section{PSF  modeling}
\label{PSF} 
The Point  Spread Function (PSF) is  modeled by fitting  the images of
sure stars  (Sec.~\ref{catalog}) with a sum  of two-dimensional Moffat
functions.  In order to account  for PSF asymmetries, the isophotes of
each  Moffat  function  are  described  by ellipses,  whose  shape  is
modulated with  a sin/cos angular  expansion, similar to  that adopted
for  describing  deviations  of  the  isophotal  shape  of  early-type
galaxies  from  pure ellipses  (see  e.g.  \citealt{Bender:87}).   The
number  of fitted  stars  is given  by  the lesser  of  the number  of
available  sure stars and  the 2DPHOT  input parameter  $NSMAX$.  The
value of  $NSMAX$ is  chosen as a  compromise between  the computation
time  for the fitting  algorithm and  the accuracy  of the  PSF model.
Increasing  $NSMAX$ yields  more accurate  PSF models  at the  cost of
larger computational  times.  Usually, values of $NSMAX$  in the range
of 3 to 5 give reliable results\footnote{\footnotesize Processing several images, we
found that increasing  the value of $NSMAX$ to more  than 5 stars does
not  significantly change  the  output of  2DPHOT.}.   To account  for
possible  spatial  variations  of  the  PSF across  the  chip,  2DPHOT
provides two  PSF modeling options.  In  the first case,  a global PSF
model is  obtained by simultaneously fitting $NSMAX$ stars 
randomly extracted  from the entire list  of sure stars.   As a second
option, 2DPHOT can construct a two-dimensional grid on the input image
and  derive a  PSF  model  independently for  each  cell, by  randomly
selecting up  to $NSMAX$  stars among the  available sure  stars.  PSF
models are only  derived for cells including at  least two sure stars.
The cell  size has to be  provided through the  2DPHOT input parameter
$NSIZE$.  2DPHOT associates  to the PSF model of  each cell the median
values of  the x  and y coordinates  of the corresponding  fitted sure
stars    and   the   two-dimensional    modeling   of    each   galaxy
(Sec.~\ref{2DFIT}) is  performed by using  its closest PSF  model.  In
order to avoid a discretely varying  PSF across the chip, the user can
also  choose to  adopt a  locally  interpolated PSF  model.  For  each
galaxy, 2DPHOT selects  the PSF models of the  cells around the galaxy
itself, and performs a  bi-linear interpolation of the selected models
at  the galaxy  position. Since  there is  a strong  correlation between the
fitting parameters of each PSF model, 2DPHOT does not derive the local
PSF model  by interpolating  each single fitting  parameter.  Instead,
the interpolation is performed independently for each pixel of the PSF
models, by interpolating the corresponding intensity values.

\begin{figure}[!]
\begin{center}
\scalebox{0.35}{\includegraphics{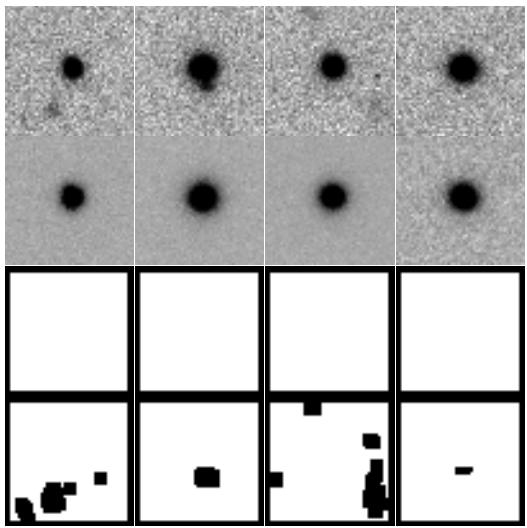}}
\end{center}
\caption[]{\footnotesize Examples of the clipping procedure applied to
four stellar images.  Each column  corresponds to a different star.  From
top to bottom,  the star stamps, the corresponding  median images, the
original mask images  and the  updated mask images  are shown.  The  four stars
have  been selected  because  S-Extractor fails  to  detect the  faint
sources around  them, and therefore the corresponding  mask images are
blank.   The  2DPHOT  clipping  procedure  detects  the  missed  faint
sources, and  masks them in the  updated mask images.  The star stamps and
median images use the same intensity scale.
\label{STARCLIP} }
\end{figure}

\begin{figure}
\begin{center}
\scalebox{0.35}{\includegraphics{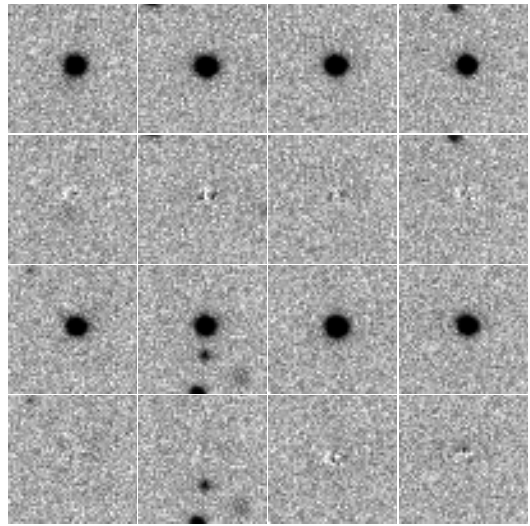}}
\end{center}
\caption[]{\footnotesize Examples of  PSF modeling.  Eight star images
  have   been  fitted  simultaneously   using   a  sum   of  three
  two-dimensional Moffat functions (see  the text). For each star, two
  panels are shown, with the  upper panel plotting the stamp image and
  the   lower  panel   showing   the  fitting   residuals.
  ~\label{PSF_FIT} }
\end{figure}

Prior  to fitting  the PSF,  2DPHOT  applies a  clipping procedure  to
remove stars that  might be contaminated by nearby  objects.  For each
star, all of the other sure stars are co-registered to the same center
coordinates  and  median stacked.   An  rms  image  is constructed  by
estimating, at  each position, the  standard deviation of  the stacked
pixels. The mask images of the sure stars are then updated by flagging
all  the  pixels  which  deviate  by  more  than  5$\sigma$  from  the
corresponding  median images.  If  the fraction  of flagged  pixels is
larger than  $20\%$ of  the total  mask image area,  the sure  star is
considered to be strongly contaminated and it is excluded from the PSF
fitting.  This procedure allows faint sources which may not have been
detected by S-Extractor to be masked, and to exclude objects which
are misclassified  or blended with nearby  sources.  Some examples
of   the  clipping   and   mask  update   algorithms   are  shown   in
Fig.~\ref{STARCLIP}, while Fig.~\ref{PSF_FIT}  plots an example of the PSF
modeling  results.  The latter  figure  is  automatically produced  by
2DPHOT.

\section{Coarse 2D fitting} 
\label{INI2DF}
2DPHOT produces an initial estimate of structural parameters for all
objects in the input image using a discrete, coarse two-dimensional
fitting algorithm (INI2DF).  For each object, a set of PSF-convolved
S\'{e}rsic models is constructed by varying the effective radius $r_{\rm
e}$, the total magnitude $m_{\rm T}$, and the S\'{e}rsic index $n$.
`Geometric' parameters, such as the center coordinates, the axis ratio
and the position angle of the models are estimated by fitting the
object image with a single 2D Moffat function, and are kept fixed
during the coarse fitting.  The Moffat fit is performed by excluding
the inner part of the object, which is strongly affected by seeing.
The local background value is also not changed in the fit and is
obtained from the keyword $M\_BK$ in the mask image header (see
Sec.~\ref{stamps}).  INI2DF changes the effective radius of the S\'{e}rsic
model using an adaptive grid of 10 values computed on the basis of
both the pixel scale and the seeing FWHM of the image.  Four different
values are considered for the S\'{e}rsic index parameter, $ n= \{1, 3, 5,
7\}$, while the total magnitude can take the values $m_{\rm T} =
m_{\rm K}, m_{\rm K}-0.2, m_{\rm K}-0.4$, where $m_{\rm K}$ is the
Kron magnitude of the source (S-Extractor $MAGAUTO$).  We point out
that the grids of $r_{\rm e}$, $m_{\rm T}$ and $n$ values have been
empirically chosen by analyzing several images with a wide range of
characteristics (e.g.  optical and near-infrared data as well as
ground-based and HST images).  We found that further increasing the
grid size does not change significantly the 2DPHOT results.  With
these sizes for the $r_{\rm e}$, $m_{\rm T}$ and $n$ grids, INI2DF
produces a total of $120$ discrete models, each of which is compared
to the object image by computing the corresponding $\chi^2$ value.
The coarse structural parameters are given by the parameters of the
model with lowest $\chi^2$.

\begin{figure}
\begin{center}
\scalebox{0.41}{\includegraphics{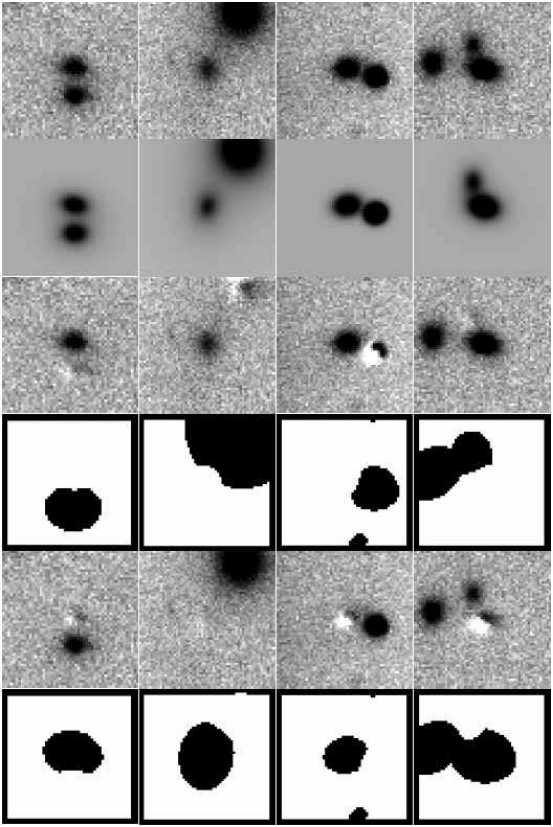}}
\end{center}
\caption[]{\footnotesize Coarse  fitting of double  objects. Panels in
  each column  of the  figure correspond to  a different  stamp image.
  From the  top row  down, the  first and second  rows show  the stamp
  image  and the  corresponding  Moffat model, obtained  by
  simultaneously fitting two single  Moffat functions. The third panel
  shows the  subtraction of one  Moffat function from the  stamp.  The
  corresponding updated mask  image is shown in the  fourth panel. The
  lowest  two  panels  show   the  Moffat  subtracted  image  and  the
  corresponding  updated  mask image  for  the  second  object in  the
  stamp. \label{MULTI} }
\end{figure}

In  the case  that, for  a  given stamp,  several objects  have to  be
treated simultaneously (see Sec.~\ref{stamps}), the above procedure is
modified  as follows.   A simultaneous  fit  is performed  by using  a
single  two-dimensional Moffat  function for  each object.   Then, for
each overlapping  object, the others  are subtracted using  the fitted
Moffat models.   A corresponding updated mask image  is also produced,
by  flagging all pixels  for which  the sum  of the  subtracted Moffat
models exceeds the local background standard deviation ($S\_BK$ in the
mask  image  header,  see   Sec.~\ref{masks})  by  $>50  \%$.   Coarse
structural parameters are then obtained  by fitting each object in the
stamp as a single source,  applying the same procedure outlined above.
Some examples of this procedure  are shown in Fig.~\ref{MULTI}. We see
that there are some cases where the single Moffat models do not result
in accurate subtraction of overlapping sources. Nevertheless, we found
that  the  above  approach  allows reliable  estimation  of  structural
parameters, with the great  advantage of significantly
reduced computational times compared  to an approach where overlapping
galaxy models are fitted simultaneously (see also Sec.~\ref{2DFIT}).

\section{Identification of stars in the 2DPHOT package} 
\label{SGCLAS}
The classification of stars and galaxies is one of the most
challenging issue in the analysis of astronomical images, and there is
no method that works in all scenarios as the optimum classifier.  In
the current version, 2DPHOT adopts a simple method of star/galaxy
(hereafter $S/G$) separation, which is based on both the S-Extractor
and the coarse structural parameters estimated by the INI2DF procedure
(Sec.~\ref{INI2DF}).  The parameters which are used for $S/G$
classification have been chosen on the basis of Monte-Carlo
simulations as detailed in Sec.~\ref{whyparam}, while the $S/G$
classification algorithm is described in Sec.~\ref{SGrules}.  In the
future, we plan to implement more complex classification techniques
(such as wavelet approaches), and provide a detailed comparison of
their performance.  Since there is no method that correctly
classifies all sources in a given image, particularly at the faintest
flux levels, it is crucial that every classification framework provide
an estimate of contamination due to misclassified sources as a
function of the S/N ratio.  As described in Sec.~\ref{SGrules}, 2DPHOT
accurately estimates such contamination using simulated stars and
galaxies added to the input processed frame.

\subsection{Reliable parameters to identify stellar sources}
\label{whyparam}

We adopt a two-step procedure to  establish   useful  parameters  for  star/galaxy
separation. First, we look for reliable
classifiers of  point-like sources,  i.e. 2DPHOT output parameters
whose values for stellar sources lie in a
narrow region of parameter space over wide ranges of the S/N ratio,
seeing,  and sampling  characteristics of the images.   Then, we
analyze  the  ability  of  such  classifiers  to  separate  stars  and
galaxies by  examining the distribution of  values they  assume for
both  kinds  of objects.   To  address  the  first point,  we  created
simulated  CCD images,  each  with a  random  spatial distribution  of
stars.   The  simulations were  generated  using  the  same pixel  scale
($0.369''/pix$),  image  size ($2048\times2048$  pixels), and
the noise properties as  the $r$-band images of the Palomar Abell
Cluster Survey  (hereafter PACS, see Gal  et al.  2000).
The PACS data have been extensively processed from the authors through
the 2DPHOT package in order  to investigate the effects of environment
on internal color gradients of  early-type galaxies (see La Barbera et
al.~2005).

Stellar images were  simulated using both the  Gaussian profile and
the Moffat law:
\begin{equation}
 P(r)=    \left[    1+    \left(   \frac{r}{r_{\rm    c}}    \right)^2
 \right]^{-\beta},
\end{equation}
where $P(r)$  is the surface brightness  of the star as  a function of
the distance $r$ to its center,  $\beta$ is the  shape parameter of
the profile, and $r_{\rm c}$ is  the Moffat scale radius,
which is  related to the  FWHM by $FWHM=2
r_{\rm c}  (2^{1/\beta}-1)^{0.5}$. For the Moffat  fits, we
set $\beta=3$, which  is the mean value for stellar images
in the PACS,  and we varied the $FWHM$ from  one star to 
another within each  simulated image according  to a  normal deviate
with central value $<FWHM>$ and width $\sigma_{FWHM}$.  Four simulated
fields  were created, labeled  F1, F2,  F3,  and F4.  The  main
simulation  parameters are  summarized in  Table~\ref{PARSIM}.
For each field, we randomly created $N=500$ stars, and we set the
parameters $<FWHM>$ and $\sigma_{FWHM}$ as follows.  For field F1,
both the $<FWHM>$ and $\sigma_{FWHM}$ are set to the median values
measured from the $r$-band PACS images.  Fields F2 and F3 simulate
observations with worse seeing conditions.  F2 has the same $<FWHM>$
as F1 while $\sigma_{FWHM}$ is doubled, mimicking the case of large
scatter in the seeing $FWHM$ across the image.  Field F3 has the same
$\sigma_{FWHM}$ as field F1 but higher $<FWHM>$, corresponding to
either observations taken in worse mean seeing or data with better PSF
sampling.  Finally, for Field F4, we used the same $<FWHM>$ and
$\sigma_{FWHM}$ values as F1, but stellar images were created
with Gaussian profiles.  We note that the above simulated images
span all the possible cases that have been found when processing the
PACS images, and because of their wide range of seeing parameters,
they also reproduce the seeing properties of a variety of ground-based
images.


\begin{figure}
\begin{center}
\scalebox{0.35}{\includegraphics{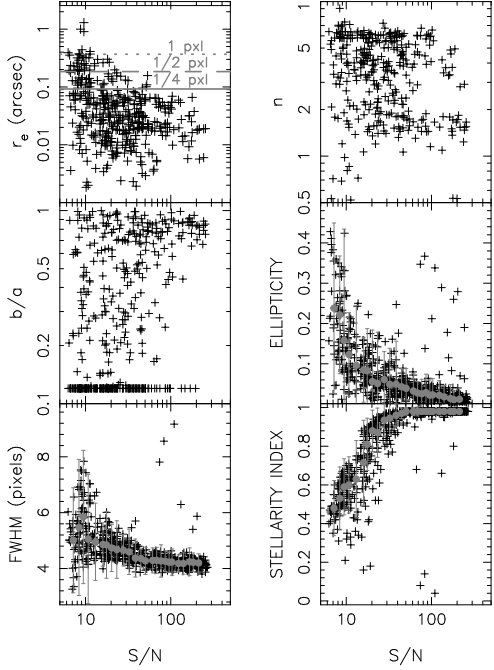}}
\end{center}
\caption[]{\footnotesize Measured  parameters for sources in
  field F1 as  a function of their S/N  ratio.  The quantities $r_{\rm
  e}$, $b/a$, and $n$ are the effective radius, axis ratio, and S\'{e}rsic
  index  obtained  from  2DPHOT  through  the  two-dimensional  fitting
  procedure (see Sec.~\ref{2DFIT}). The other parameters (ELLIPTICITY,
  FWHM, and STELLARITY INDEX)  are those measured by S-Extractor. As
  shown in the plot, the  solid, dashed and short-dashed grey lines in
  the upper-left panel mark the values of $r_{\rm e}$ corresponding to
  different pixel fractions. Grey circles and corresponding error bars
  in  the ELLIPTICITY,  FWHM, and  STELLARITY INDEX  panels  have been
  obtained  by binning the  data with  respect to  the S/N  ratio, and
  correspond to the mean and 1$\sigma$ interval in each bin.
\label{F1} }
\end{figure}

\begin{figure}
\begin{center}
\scalebox{0.35}{\includegraphics{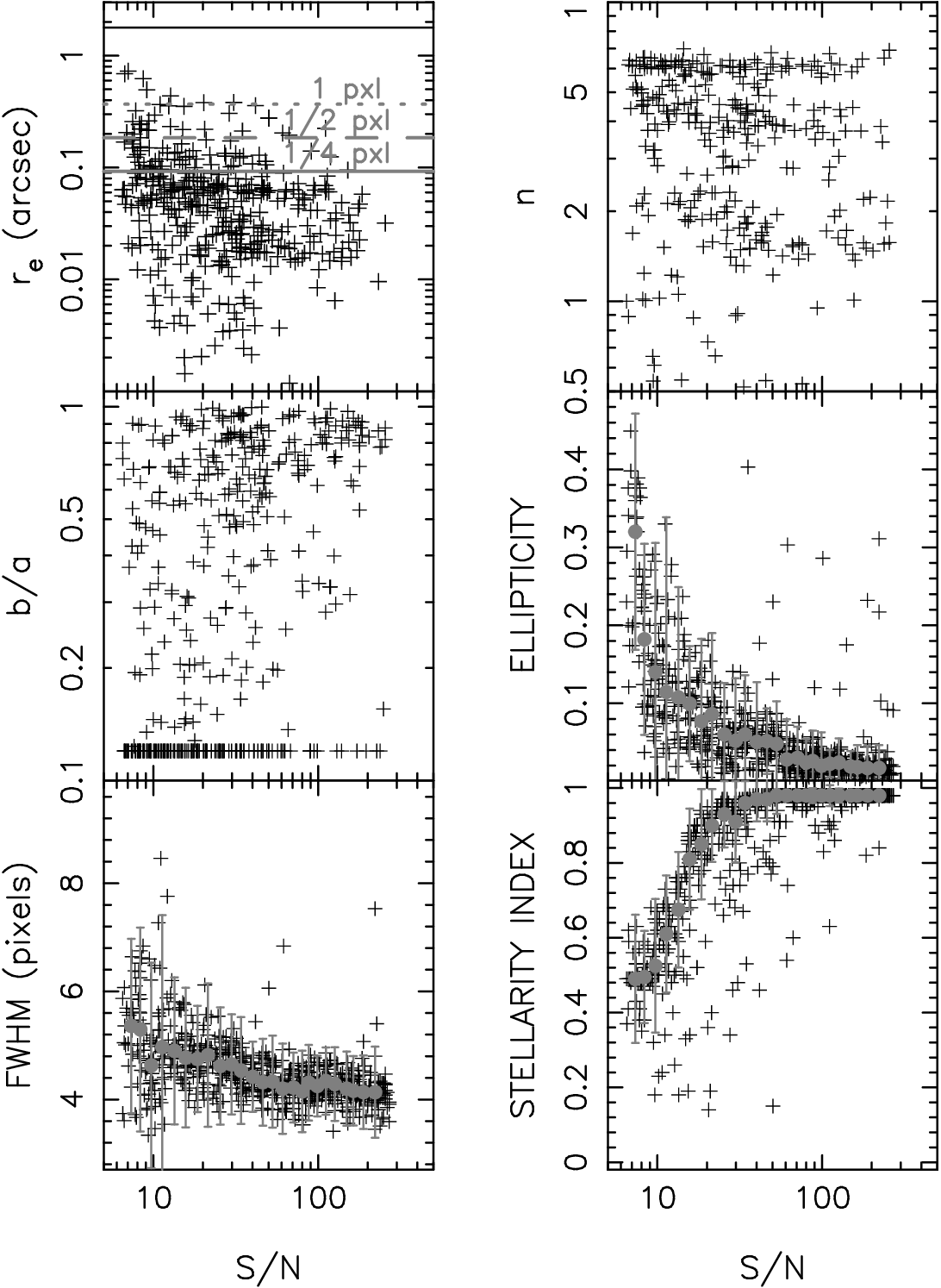}}
\end{center}
\caption[]{\footnotesize Same as Fig.~\ref{F1} but for Field F2. \label{F2} }
\end{figure}

\begin{figure}
\begin{center}
\scalebox{0.35}{\includegraphics{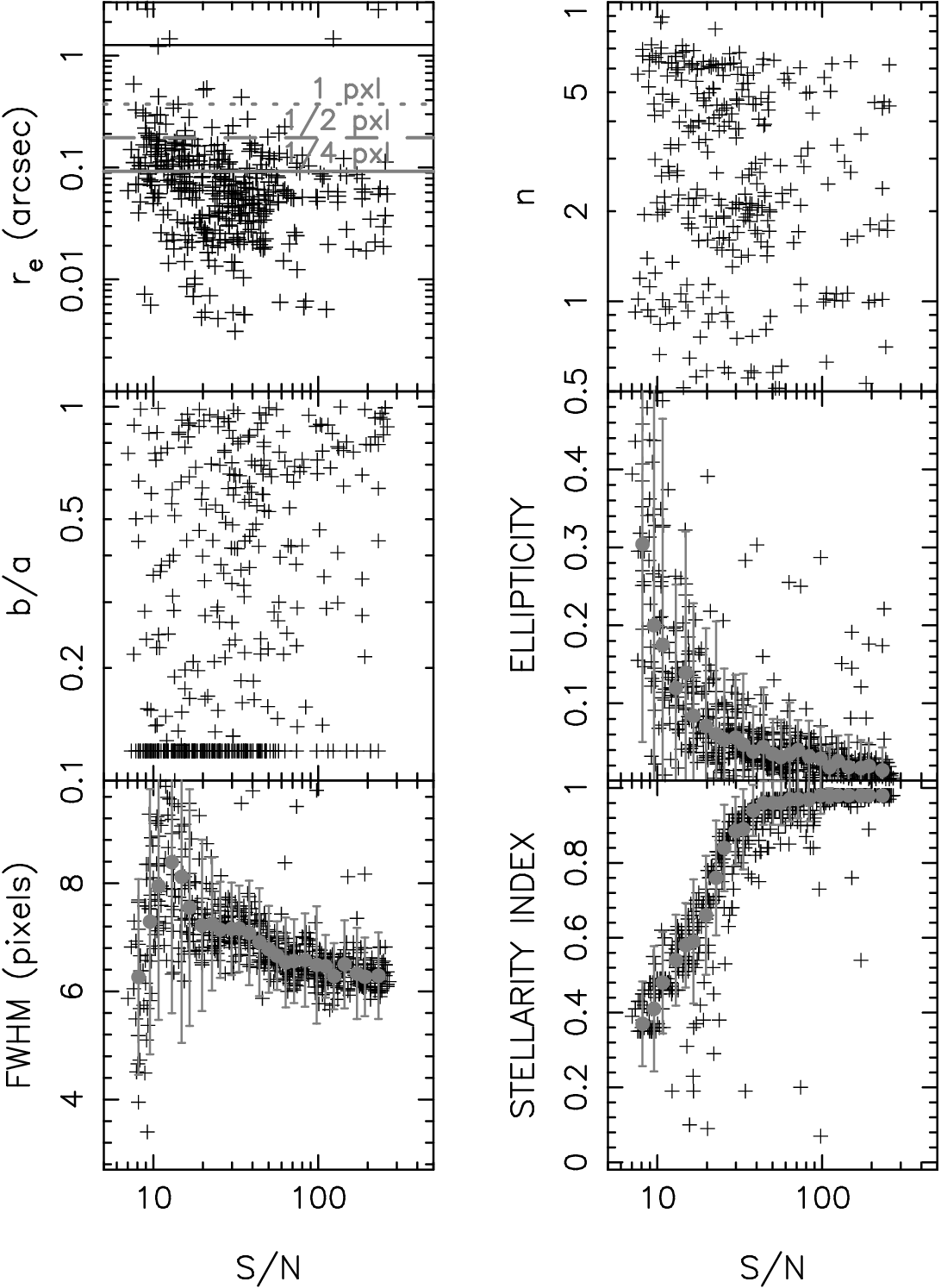}}
\end{center}
\caption[]{\footnotesize  Same as Fig.~\ref{F1} but for Field F3. \label{F3} }
\end{figure}

\begin{figure}
\begin{center}
\scalebox{0.35}{\includegraphics{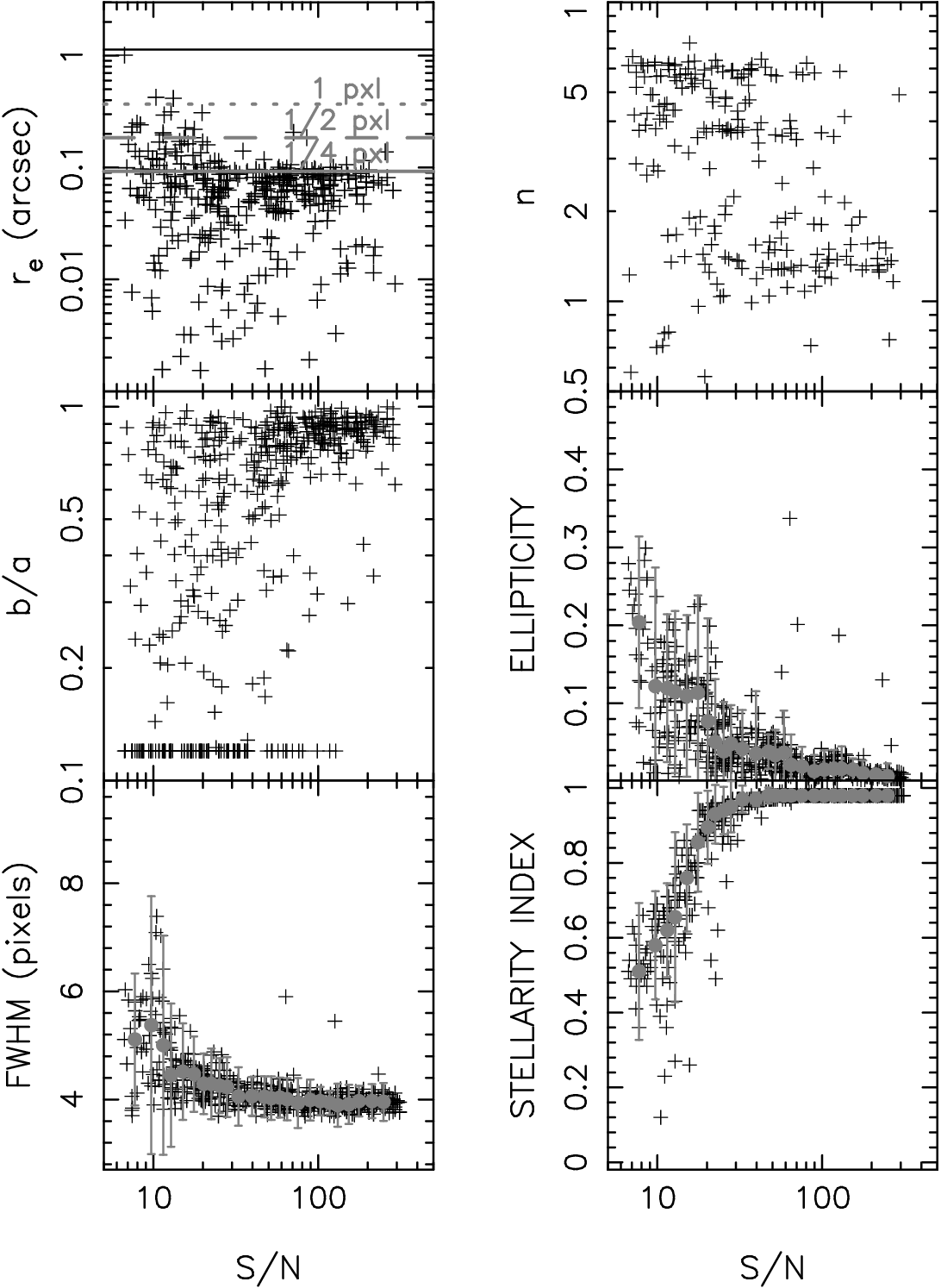}}
\end{center}
\caption[]{\footnotesize  Same as Fig.~\ref{F1} but for Field F4. \label{F4} }
\end{figure}

Catalogs of the simulated stellar  fields were generated as described in
Sec.~\ref{catalog}.   For each  field, all  detected  sources were
fit with  PSF convolved S\'{e}rsic  models, following the  procedure
described in  Secs.~3--5 and running the  final two-dimensional fitting
program  (see Sec.~\ref  {2DFIT}).  Figs.~\ref{F1},~\ref{F2},~\ref{F3}
and~\ref{F4} plot  the S\'{e}rsic  parameters, i.e.  the  effective radius
$r_e$, the S\'{e}rsic index $n$, and  the axis ratio $b/a$, as well as the
the ELLIPTICITY,  FWHM and stellarity index  (hereafter SI) parameters
from S-Extractor as a function of the S/N ratio of sources in fields F1,
F2,  F3, and  F4, respectively.   The S/N  ratio was  computed  as the
inverse of  the uncertainty on  the S-Extractor Kron  magnitude.  From
Figs.~\ref{F1},~\ref{F2},~\ref{F3} and~\ref{F4}, we draw the following
conclusions:
\begin{description}
\item[i)] The S\'{e}rsic  index and the $b/a$ parameters  are not reliable
  classifiers.  The scatter in these quantities is large compared
  to the range of values they can assume.
\item[ii)] The effective radius is a reliable classifier, in the sense
  that its  values are always  well limited to  a given region  of the
  corresponding parameter  space. Whatever the  seeing conditions are,
  the effective radius of stars  is always smaller than $\sim 1$ pixel,
  and for $S/N>30$, the values  of $r_{\rm e}$ are always smaller than
  $\sim 0.5$ pixel.
\item[iii)] As one would expect~\citep{BeA:96}, the SI parameter of
  S-Extractor is a reliable classifier. Its values can range from 0 to
  1, but  for the simulated stars  with $S/N >20$, the  values of $SI$
  are always larger than $\sim0.7$.
\item[iv)]  The FWHM  and  ELLIPTICITY parameters  are good  potential
  classifiers  as well,  although  the values  of  FWHM are  obviously
  strongly  dependent on  the seeing  characteristics of  the analyzed
  image.   Generally, we  find  that using  the  FWHM and  ELLIPTICITY
  parameters  does not  lead  to any  significant  improvement in 
  star/galaxy separation,  and thus we  elected not  to use
  these parameters.
\end{description}
The reliability  of the above star/galaxy separation  scheme, based on
the $SI$--$r_{e}$ diagram,  will be further  addressed in  Section 13,
where  we will assess  contamination and  completeness as  measured by
2DPHOT.
\begin{table*}
\caption[]{Parameters of simulated  stellar fields.  Cols.~2,~3 and~4
give the  $FWHM$, $\sigma_{FWHM} /  FWHM$ and $\beta$  parameters (see
text). The gain, zero-point and read-out noise are in Cols.~5,~6 and 7,
respectively. In  the case  of field 4,  stellar images  have Gaussian
profiles.\label{PARSIM}}
\hspace{2.3cm}
\begin{tabular}{|c|cccccc|}
\hline
$Field  \#$ & $FWHM$  & $\frac{\sigma_{\rm  FWHM}}{FWHM}$ &  $\beta$ &
        $gain$ & $zpoint$ & $rnoise$ \\  & (pxls) & $\%$ & & $e^-/ADU$
        & & $e^-$ \\ 
\hline
1 & 4 & 3 & 3 & 1.62 & 30.75 & 6.3 \\ 
2 & 4 & 3 & 3 & 1.62 & 30.75  & 6.3 \\ 
3 & 6 & 6 & 3  & 1.62 & 30.75 & 6.3 \\ 
4 & 4 & 3 & Gaussian  & 1.62 & 30.75 & 6.3 \\ \hline

\end{tabular}
\end{table*}

\begin{figure}[!]
\begin{center}
\scalebox{0.4}{\includegraphics{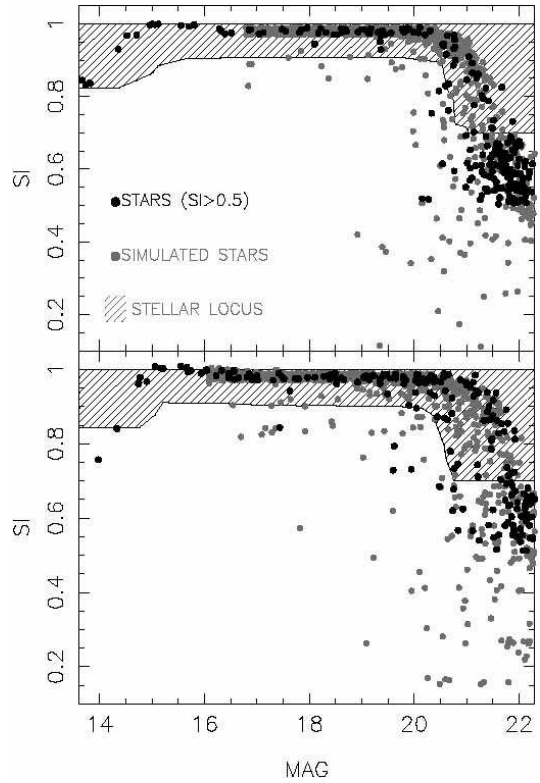}}
\end{center}
\caption[]{\footnotesize Definition of the stellar locus in the 2DPHOT
  package (see text). The plot shows the stellar index versus Kron
  magnitude diagram  for two $g$-band  images from PACS.   Grey points
  are  the simulated  stars added to  each  field by
  2DPHOT,  while black  circles show  the objects  with  stellar index
  larger than 0.5. The hatched area marks the region used
  to select star candidates.\label{STARLOCUS} }
\end{figure}

\begin{figure}
\begin{center}
\scalebox{0.4}{\includegraphics{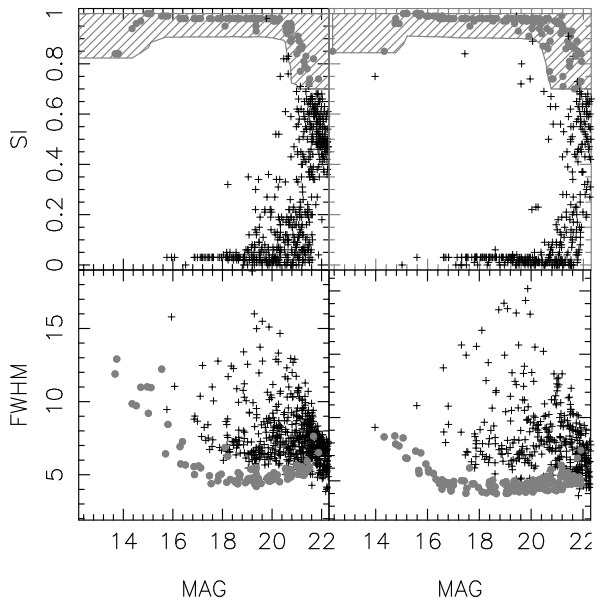}}
\end{center}
\caption[]{\footnotesize Selection of stars  in 2DPHOT for
  two $g$-band  images of the PACS.   The upper panels  show the stellar
  index versus  Kron magnitude diagrams, with the  hatched region
  showing  the stellar  locus (see  text).   Grey circles  are the
  selected stars, while black  crosses plot all the remaining sources.
  Lower panels show the corresponding FWHM versus Kron magnitude plots
  for the same PACS images.~\label{STARDEF} }
\end{figure}

\subsection{Final rules for identifying stellar sources}
\label{SGrules}
After INI2DF  parameters have  been derived for  all the sources  in a
given field,  2PHOT performs  $S/G$ separation.  Simulated  images are
constructed by adding a spatially  random distribution of stars to the
input image.  The surface density of  stars is chosen so that 50 stars
are added to an input image area of 2000$\times$2000 pixels, while the
number of simulations is such that  we have a total of 2000 artificial
stars.  Stars are created from the actual PSF model, with photon noise
added  based on the  GAIN provided  in the  {\it default.sex}  file of
S-Extractor.   The  magnitude of  each  artificial  star is  extracted
according to a uniform random  $S/N$ ratio distribution, with an upper
cutoff of $S/N=200$.  For each  simulation, a new catalog is generated
and  the SI  parameter is  computed for  all of  the  artificial stars
detected by S-Extractor.  2DPHOT  defines star candidates on the basis
of  the  distribution of  these  artificial  stars  in the  SI  versus
magnitude diagram.   First, artificial stars are  ordered by ascending
magnitude, and  the 50 stars with  magnitudes closest to  that of each
artificial  star are  selected.   Then, for  each  artifial star,  the
$10\%$ percentile of the distribution  of SI values ($SI_{10}$) of the
50 selected  artificial stars is  computed.  In order to  minimize the
number of  galaxies that are misclassified at  the faintest magnitudes
of  the catalog,  a minimum  cutoff of  $SI_{10}=0.7$ is  imposed (see
Sec.~\ref{SG_CONTAM} for details). As  a second option, that turns out
to be more  suitable in the case of deep  images (see Sec.~13), 2DPHOT
can  define the  star locus  by applying  the same  procedure outlined
above but replacing the $10\%$  percentile of the $SI$ distribution of
simulated stars  with the quantity $\theta- p  \sigma$, where $\theta$
and  $\sigma$  are  the  location  and  width  of  the  $SI$
distribution,  while $p$  is  a parameter input to  2DPHOT. With  a
suitable  choice of  $p$ this second  definition  allows a  narrower
stellar locus to be defined  in the $SI$ versus magnitude diagram (see
Sec.~\ref{SG_CONTAM}),  and thus  it  can be  more  suitable at  faint
magnitudes  where  galaxies  with   small  size are more likely to be 
misclassified  as  stars.   Hereafter,  unless  stated explicitly,  we  will
consider only the first definition of the star locus.

Since saturated  stars have lower $SI$ than  bright unsaturated stars,
at magnitudes  brighter than those  of artificial stars the  value of
$SI_{10}$ is  set to the minimum  $SI$ for observed  sources with $SI>
SI_{\rm min}=0.5$.  This procedure allows  even saturated stars  to be
correctly classified by 2DPHOT. The value of $SI_{\rm min}$ was chosen
empirically  based on  several processed  images where  we  found that
saturated stars always have $SI>0.5$, while the brightest galaxies all
have lower $SI$. As shown in Fig.~\ref{STARLOCUS}, star candidates are
selected using the region between  $SI=SI_{10}$ and $SI=1$ in the $SI$
versus  magnitude diagram.  Plots  like those  in Fig.~\ref{STARLOCUS}
are automatically  produced by each  run of 2DPHOT.  If  required, the
value of $SI_{\rm min}$ can be changed by the user after inspection of
the  stellar locus  plot.  As  final  rules for  $S/G$ separation,  we
define  an object  as  a candidate  star  if it  belongs  to the  star
candidate  locus and  its INI2DF  effective radius  is smaller  than 1
pixel,  with the  latter  criterion from  the  results of  Monte-Carlo
simulations  discussed  in  Sec.~\ref{whyparam}.   The locus  of  star
candidates should include most of  the point-like sources in the input
image, with the percentage  of misclassified objects increasing as the
magnitude increases.  The selection of stars through this procedure is
shown in Fig.~\ref{STARDEF}, where we consider two CCD images from the
PACS.  The  plots in Fig.~\ref{STARDEF} are  automatically produced by
2DPHOT.   As one  would expect,  at faint  magnitudes  the star/galaxy
classification becomes progressively  more uncertain. The distribution
of  artificial stars  with respect  to  the locus  of star  candidates
provides a quantitative way to estimate the magnitude (and/or) the S/N
limit  above   which  the   $S/G$  classification  is   reliable  (see
Sec.~\ref{SG_CONTAM}).

\section{Final 2D fitting}
\label{2DFIT}
\begin{figure*}
\begin{center}
\scalebox{0.75}{\includegraphics{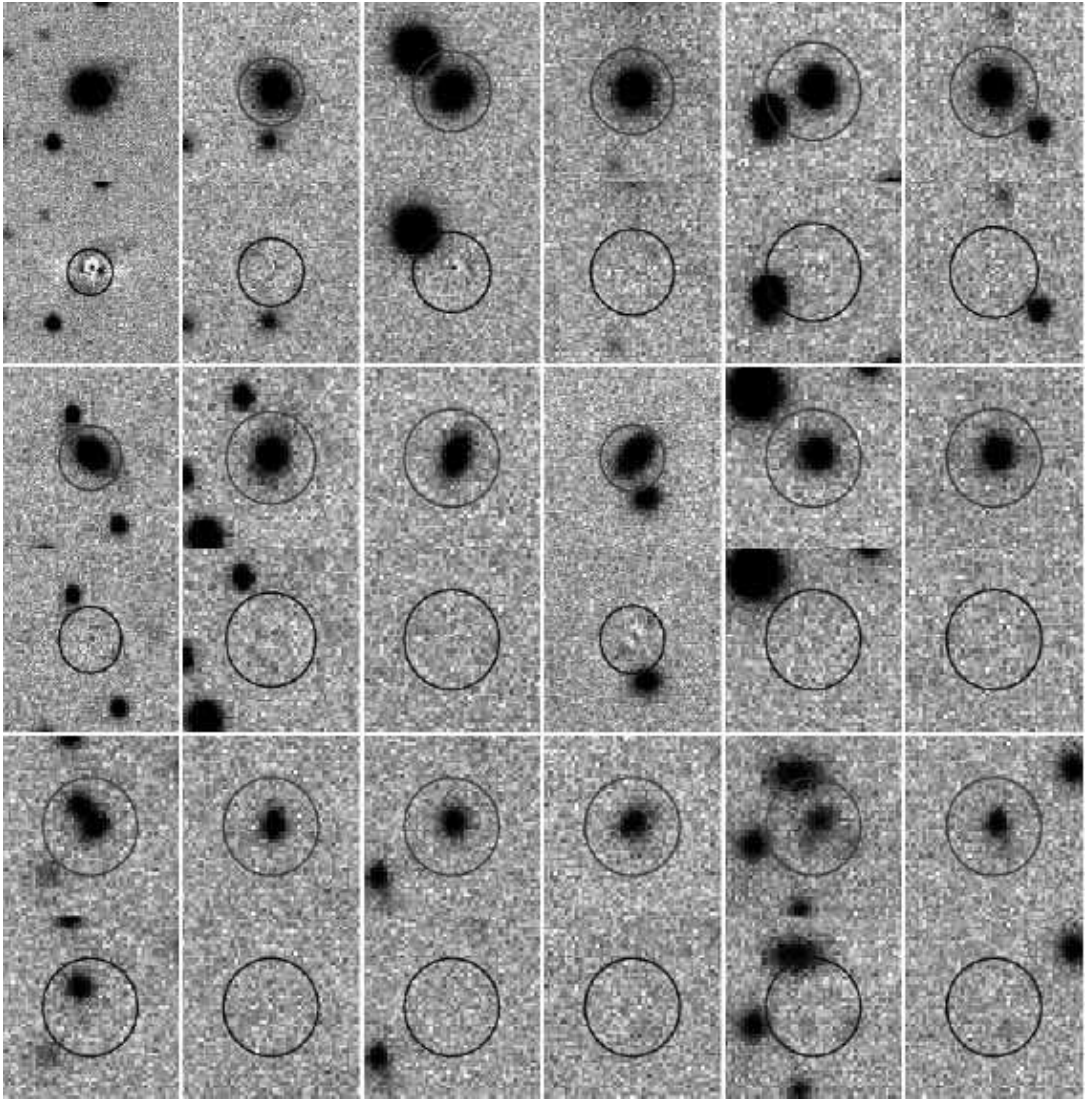}}
\end{center}
\caption[]{\footnotesize Two-dimensional fitting of galaxy stamps with
  seeing-convolved  S\'{e}rsic models.   Subpanels show the  galaxy stamps
  and the  corresponding residual images, obtained by
  subtracting the model fit from the galaxy stamp.  For each galaxy
  stamp, the fitted galaxy is marked by a grey circle of radius $2''$,
  and  the lower  subpanel shows  the residual  image, where  the same
  circle is plotted  in black.  From top to bottom  and left to right,
  galaxies  are   shown  in   order  of  decreasing   magnitude,  from
  $r\sim16.5$  for   the  upper-right  panel  to   $r\sim20$  for  the
  lower-right  panel. Images  are drawn  from  the PACS  image of  the
  cluster Abell~574 at $z \sim 0.185$.~\label{PLOT_2DFIT} }
\end{figure*}

Objects identified as galaxies through the 2DPHOT $S/G$ classification
scheme  are then  fit with  PSF  convolved S\'{e}rsic  models.  This  `final'
fitting differs  from that of Sec.~\ref{INI2DF} since  a full $\chi^2$
minimization algorithm is adopted, without using any discrete (coarse)
grid  of  reference convolved  models  (as  for  INI2DF), providing  a
precise  estimate  of structural  parameters  at  the  cost of  longer
computation  times\footnote{\footnotesize The  CPU   time  required  for  the  final
two-dimensional fitting is 4-5 times longer than for the coarse fit.}.
The $\chi^2$ minimization is performed through the Levenberg-Marquardt
algorithm, assigning zero weight to all the flagged pixels in the mask
image.  The 2D  fitting routine adopted in the  2DPHOT package is also
described in \citet{LaB:02}, where several tests of its accuracy
have  been performed.   The  initial conditions  for the  optimization
routine are set  to the output values of INI2DF,  which are on average
quite close to the best fitting final parameters. This largely reduces
the  well known issue  of spurious  convergence that  can characterize
strongly  non-linear optimization problems.   The case  of overlapping
objects is  treated with  an analogous approach  to that  described in
Sec.~\ref{INI2DF} for  the coarse fit.  Instead of  using the multiple
single  Moffat  fits  described  in  Sec.~\ref{INI2DF},  2DPHOT  takes
advantage  of the  INI2DF  best  fitting models  to  reduce the  final
fitting  of overlapping objects  to that  of separate  single sources.
For each blended galaxy,  the overlapping objects are subtracted using
the INI2DF  models and  the mask image  is correspondingly  updated as
described  in  Sec.~\ref{INI2DF}.  Although  a  suitable treatment  of
overlapping galaxies would require  a simultaneous fit to be performed
(see  e.g.~\citealt{vanDokkum:96}), reducing  the problem  to  that of
fitting single sources greatly decreases computation times.  Comparing
both  approaches,  we verified  that  the  2DPHOT  procedure does  not
produce any  significant change in the final  structural parameters of
{\it multiple} objects.  Some  examples of two-dimensional fitting are
shown  in  Fig.~\ref{PLOT_2DFIT}.   The   plots  in  this  figure  are
automatically generated by 2DPHOT.

\section{Isophotal analysis}
\label{SPHOT}
To  analyze  the isophotal  properties  of galaxies,  2DPHOT
performs  an  elliptical fit  of  galaxy  isophotes  and measures  the
deviations of  such isophotes from  purely elliptical shapes.   Details on
how the  package performs these  tasks are given  in Sec.~\ref{ISFIT}.
The  isophotal fit  allows the  radial surface  brightness  profile of
galaxies   and    stars   to   be   extracted.     As   described   in
Sec.~\ref{SBPROF},  the  package  uses  these brightness  profiles  to
obtain a  further estimate of the galaxy  structural parameters, hence
providing  an  independent  estimate  of these  parameters  than  that
obtained    with   the    full   two-dimensional    fitting   approach
(Sec.~\ref{2DFIT}). The  isophotal analysis is  also used to  extract a
growth  curve for each  galaxy's  aperture  magnitude. The  aperture
magnitudes are computed and  corrected for seeing effects as described
in Sec.~\ref{GROWTH}.

\subsection{Isophotal fitting}
\label{ISFIT}
\begin{figure*}
\begin{center}
\scalebox{0.4}{\includegraphics{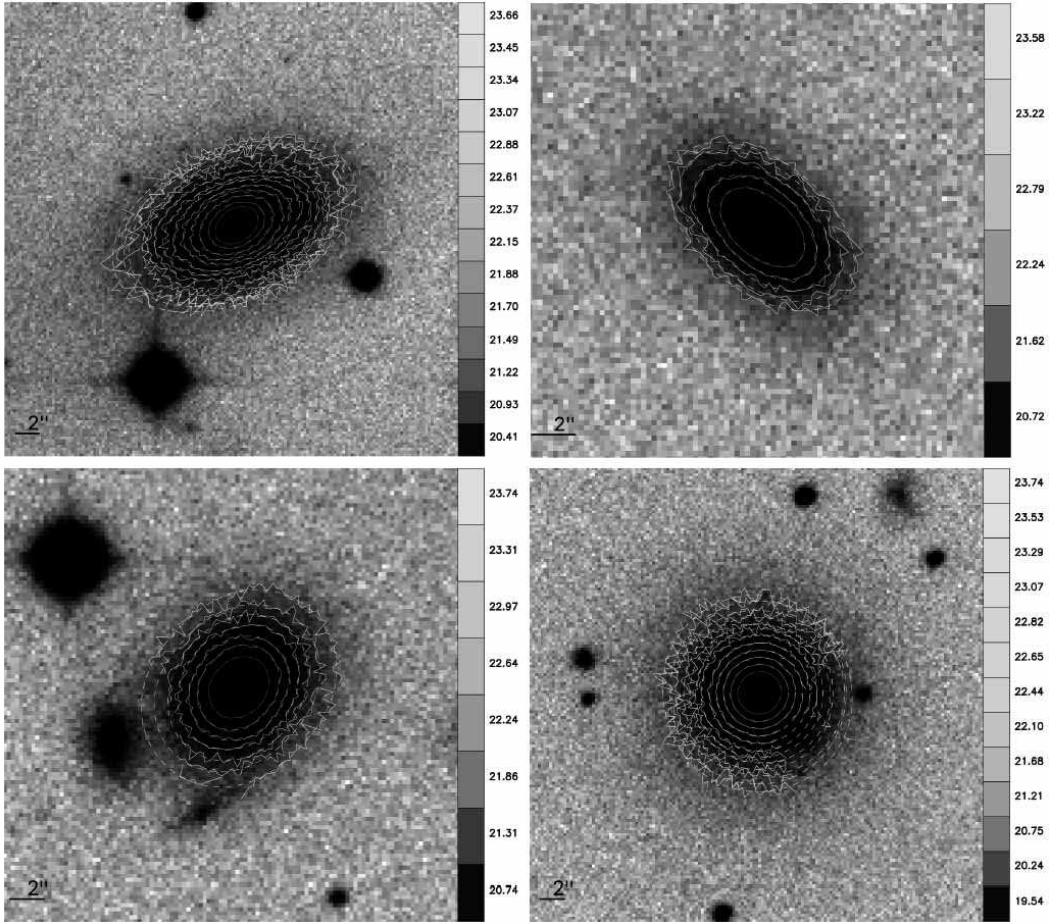}}
\end{center}
\caption[]{\footnotesize Fitting of galaxy isophotes with elliptical
contours modulated by a series of sin/cos angular functions. The four
panels plot different galaxy stamps, extracted from one PACS $r$-band
image.  Solid lines are the isophotal contours, derived as described
in Sec.~\ref{ISFIT}, while the fitted ellipses are plotted as dashed
curves. The fits were performed by including only the $a_4$ term in
the sin/cos expansion.  Isophotes are plotted with different gray
levels, with the grayscale proportional to the corresponding surface
brightness value.  The relations between gray intensity and surface
brightness are shown on the grey scales at the right of each panel.
Each surface brightness value on these gray scales corresponds to a
different isophote. Surface brightness values are given in units of
$mag/arcsec^2$, and become brighter as the isophotal color changes
from white to black.  The spatial scale is shown in the lower-left
corner of each panel.  The plots are produced automatically from the
2DPHOT package.~\label{IS_PLOT} }
\end{figure*}

For the measurement of galaxy isophotes, the package first defines the
corresponding  isophotal intensity  values.  For  each stamp,  a rough
estimate  of  the  object  center  coordinates  are  obtained  as  the
intensity-weighted means of the x  and y pixel coordinates.   The mean
values are  computed in a section  of 5x5 pixels  around the intensity
peak  of  the  object.   Using  these center  coordinates,  a  set  of
concentric circles  is constructed, with  radii equally spaced  by 0.5
pixel. For each circle, the mean value of 90 intensity samples equally
spaced in polar angle is computed via cubic interpolation of the stamp
intensity values  at the  corresponding radial and  polar coordinates.
The mean  intensity values provide  an initial estimate of  the object
surface  brightness profile,  and  are used  to  derive the  isophotal
contours  of  the  object.  This  procedure  allows  us  to  construct
isophotal  contours whose equivalent  radii are  approximately equally
spaced by 0.5 pixels.  For a given isophotal intensity level, $I$, the
isophote is  defined by a  set of  x and y  pairs on the  stamp. These
isophotal samples  are defined  as follows.  For  a given  polar angle
$\theta$,  different intensity  values $I_j$  are computed  at several
radii  $r_j$  from  the  galaxy  center.  The  algorithm  selects  the
smallest radius at which the intensity brackets the value of $I$ (i.e.
$ I_j \!  \le \!  I \!  \le \!  I_{j+1}$ or $ I_{j+1} \!  \le \!  I \!
\le \!   I_{j}$).  The radius  $r$ corresponding to this  intensity is
then computed by linear interpolation of the $r_j$ values with respect
to $I_j$.   The isophotal samples  are directly computed from  $r$ and
$\theta$  by varying  $\theta$  such  that the  number  of samples  is
proportional  to the  isophote length  and by  excluding  those points
flagged in the mask file.  To exclude low signal-to-noise regions, the
isophotal computation is  stopped when the background-subtracted value
of $I$ falls below four times the background standard deviation within
the stamp.  As a default, to exclude galaxies whose isophotal contours
are overly affected by  seeing, 2DPHOT performs the isophotal analysis
only for galaxies  whose S-Extractor isophotal
radius\footnote{\footnotesize This is
defined  as  $\sqrt{ISOAREA/\pi}$,  where  $ISOAREA$  is  the  ISOAREA
parameter of S-Extractor.}  is larger than four times the seeing FWHM.
Figure~\ref{IS_PLOT}  plots  some  example  of  isophotal  analysis  for
galaxies from one PACS $r$-band image.  The  panels shown in the
plot are automatically produced by 2DPHOT.

\begin{figure}
\begin{center}
\scalebox{0.43}{\includegraphics{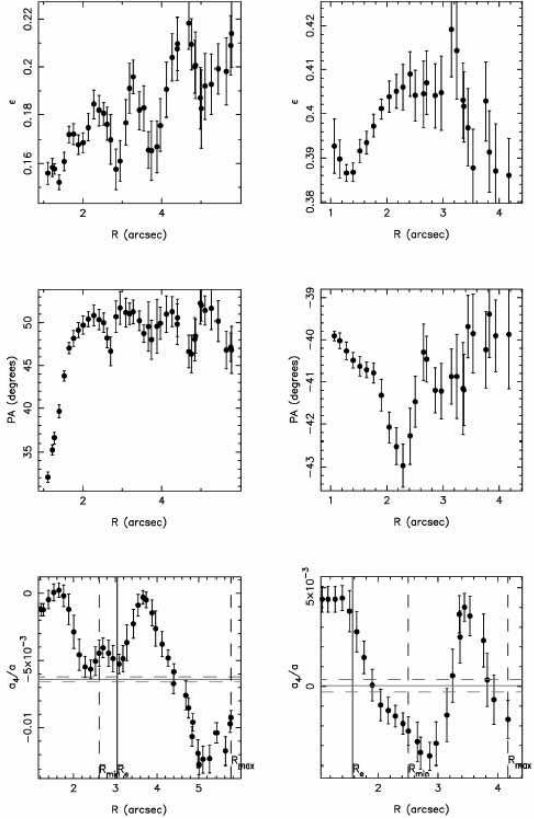}}
\end{center}
\caption[]{\footnotesize Radial  profiles of isophotal  parameters, as
derived  by the isophotal  fitting algorithm  (Sec.~\ref{ISFIT}). From
top to bottom,  the panels show the profiles  of ellipticity, position
angle of the ellipse's major axis, and $a_4$ coefficient as a function
of the  equivalent radius of  the fitted isophotes.  Left  panels show
the   profiles   of   the   galaxy   in  the   lower-left   panel   of
Fig.~\ref{IS_PLOT}, while right panels  correspond to the galaxy shown
in the  upper-right panel of  Fig.~\ref{IS_PLOT}. Error bars  mark one
sigma standard  uncertainties. In the  bottom panels, the  minimum and
maximum radii to  define the global $a_4$ value  are shown as vertical
dashed  lines. The  effective  radius of  the  galaxy is  marked by  a
vertical solid line.  The solid  horizontal gray line denotes the mean
value of  $a_4$ in  the selected radial  range, while the  dashed gray
lines   mark   the   corresponding $1\sigma$  interval.
~\label{IS_PROF} }
\end{figure}

\begin{figure}
\begin{center}
\scalebox{0.36}{\includegraphics{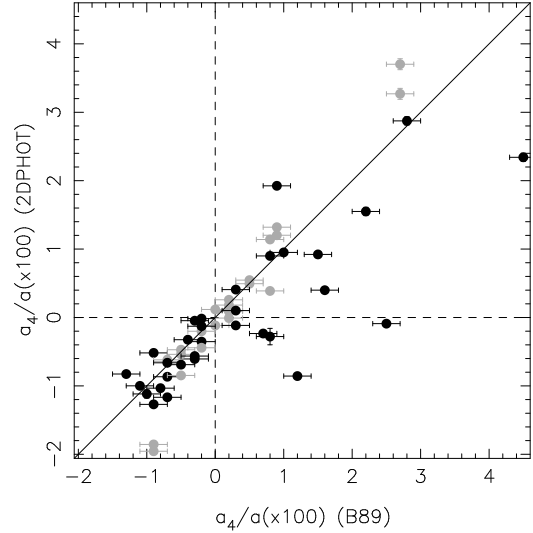}}
\end{center}
\caption[]{\footnotesize Comparison of  $a_4$ values as estimated from
 \citet{Bender:89}  (horizontal  axis)  and  from the  2DPHOT  package
 (vertical axis).  The  new $a_4$ values were obtained  by running the
 2DPHOT   package   on  r-band   images   of   the   42  galaxies   of
 \citet{Bender:89} with  available photometry  from the SDSS  DR5. The
 horizontal  error bars  mark  the typical  uncertainty  on the  $a_4$
 values  of \citet{Bender:89}  (see their  sec.2).  The  vertical bars
 denote   one   sigma  standard   uncertainties   as  estimated   from
 2DPHOT. Most of these error bars  are smaller than the symbol size in
 the plot.   From left to  right, the following galaxies  are plotted:
 NGC4261,  NGC4365,  NGC4387,   NGC5322,  NGC3605,  NGC5127,  NGC4478,
 NGC5532,  NGC3894,  NGC4551,   NGC4406,  NGC5576,  NGC4649,  NGC4374,
 NGC4472,  NGC3842,  NGC6411,   NGC4636,  NGC4489,  NGC3608,  NGC3640,
 NGC4486,  NGC5638,  NGC3379,   NGC3193,  NGC4494,  NGC5490,  NGC5831,
 NGC3613,  NGC4382,  NGC4168,   NGC5845,  NGC4125,  NGC4473,  NGC2693,
 NGC3377,  NGC4621,  NGC4550,   NGC4564,  NGC3610,  NGC4660,  NGC4251,
 NGC4570.   In   several  cases,   a  galaxy  has   repeated  SDSS
 observations.   Such  cases have  been  processed independently  by
 2DPHOT, and the corresponding values  are plotted as gray symbols in
 the figure. ~\label{BCOMP} }
\end{figure}

Galaxy  isophotes are  modeled as  described  in~\citet{Bender:87}, by
fitting  each isophote  with an  elliptical contour  modulated  by the
following sin/cos angular expansion:
\begin{equation}
 \sum a_n \cdot cos(n \theta) + b_n \cdot sin(n \theta),
\label{sincos}
\end{equation}
where $\theta$ is the polar angle, and the sum is done with respect to
the  index $n$.   For  $n \ge  3$,  the coefficients  $a_n$ and  $b_n$
describe the  deviations of the  isophotes from the  elliptical shape.
In particular, the $a_4$ term is  used to describe the boxy ($a_4 <0$)
and  disky ($a_4>0$)  isophotal  shapes of  early-type galaxies.   Each
ellipse  is characterized by  five fitting  parameters, which  are its
center coordinates, equivalent radius, ellipticity, and position angle
of the  major axis.  The  sin/cos terms which  have to be  included in
Eq.~\ref{sincos}  are  defined as  input  parameters  of 2DPHOT.   The
isophotal parameters are derived by a $\chi^2$ minimization procedure,
through a  Levenberg-Marquardt algorithm.  Examples  of isophotal fits
are  shown in Fig.~\ref{IS_PLOT}, while  Fig.~\ref{IS_PROF} shows the
radial  profiles of  isophotal  parameters  measured by  the
fitting  procedure.   All  of these  plots  are  automatically
produced by 2DPHOT.  Global values  of $a_n$ and $b_n$ are computed as
follows.  Following~\citet{Bender:87} and~\citet[hereafter
B89]{Bender:89}, only  the range  of $a_n$ and  $b_n$ profiles between  a minimum
radius $R_{min}$ and a  maximum radius $R_{max}$ is selected. $R_{min}$ is set to  four times the  seeing FWHM  of the
image, while $R_{max}$ is set  to twice the galaxy effective
radius.  The  global $a_n$ and  $b_n$  values  are then defined  as the
average    of   their    profiles   within    the    selected   radial
range. Fig.~\ref{BCOMP}  compares the $a_4$  values of B89  with those
derived by running  2DPHOT on the $r$-band images of  42 galaxies from B89
with  available photometry  from  the Sloan  Digital  Sky Survey  Data
Release 5 (SDSS DR5). Ten out of the 42 galaxies have been observed
multiple times in the SDSS, and we used  these repeated observations
to check the reliability of  the $a_4$ values.  Looking at the figure,
we  see  that there  is  good agreement  between  the  two sets  of
measurements. Moreover, there is excellent agreement among repeated
$a_4$ measurements.   We note that  2DPHOT measures global $a_n$  and $b_n$
 values   somewhat   differently than \citet{Bender:89}, where either the peak  values or the values of $a_n$ and  $b_n$ at one
effective radius  were considered.  Using  the mean
values has  the advantage of  producing more robust  estimates, reducing
the effects of possible spurious peaks in the $a_n$ and $b_n$ profiles
that can arise  from noise fluctuations. Furthermore, as  shown above, the
two methods give, on average, fully consistent results.

\subsection{Measuring surface brightness radial profiles}
\label{SBPROF}
\begin{figure}
\begin{center}
\scalebox{0.2}{\includegraphics{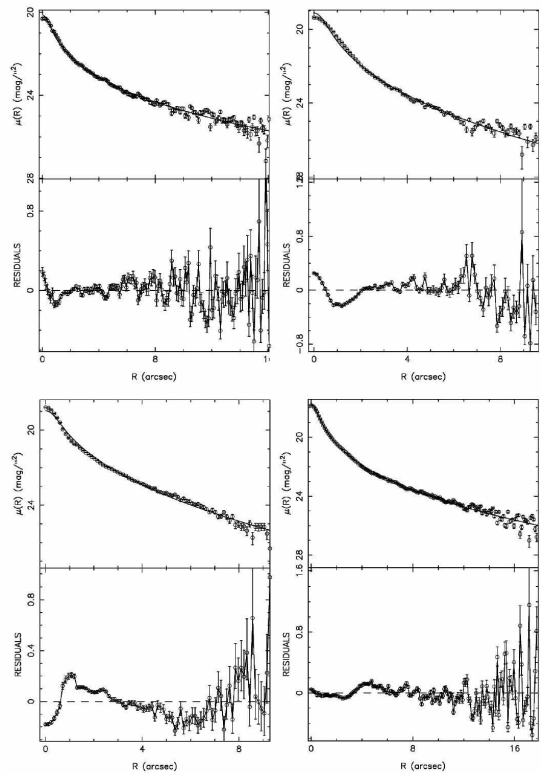}}
\end{center}
\caption[]{\footnotesize Surface brightness profiles of the same
galaxies as in Fig.~\ref{ISFIT}. In the upper plot of each panel, the
surface brightnesses computed over different elliptical contours are
plotted as a function of the ellipses' equivalent radii.  The surface
brightness values have been sky subtracted as described in the text.
The error bars denote $1\sigma$ uncertainties, computed by adding in
quadrature the standard deviation of the intensity values in each
ellipse with the uncertainty in the background estimate.  The solid
line is the best-fitting one dimensional S\'{e}rsic model.  The lower plot
of each panel shows the residuals, in units of $mag/arcsec^2$,
obtained after subtracting the model from the data. The four panels
correspond to the same galaxies as in Fig.~\ref{IS_PLOT}.
~\label{1DPROF} }
\end{figure}

For all galaxies with final 2D fitting parameters, 2DPHOT extracts a
one dimensional surface brightness profile.  Four galaxy isophotes,
corresponding to intensity values of ${4, 6, 8,}$ and 10 background
standard deviations over the background level are computed, and are
fitted by elliptical contours, as described in Sec.~\ref{ISFIT}.  The
values of center coordinates, axis ratio, and position angle of the
fitted ellipses are averaged, and are used to construct several
concentric ellipses on the galaxy stamp, with their equivalent radii
equally spaced by 0.5 pixel.  The one dimensional surface brightness
profile is then obtained as described in Sec.~\ref{ISFIT}, by
computing the mean intensity value in each ellipse as a function of
the ellipse equivalent radius.  The brightness profile is
sky-subtracted by applying a similar procedure to that described
in~\citet{Jorgensen:95}. The outermost part of the surface brightness
profile intensities is fit with a power law, $\alpha \cdot r^{-\beta}
+ bg$, where the $\alpha$ and $\beta$ parameters as well as the local
background value, $bg$, are estimated by a $\chi^2$ minimization
procedure.  The outermost part of the profile is defined as that
radial range where the mean isophotal intensity minus an approximated
median background falls below twice the background standard deviation.
Some examples of surface brightness profiles for the same galaxies as
in Fig.~\ref{IS_PLOT} are shown in Fig.~\ref{1DPROF}.  The profiles
are used to obtain a further estimate of galaxy structural parameters
independent of the 2D fitting approach.  2DPHOT follows the procedure
described by \citet[hereafter BPZ82]{Bendinelli:82}.  In this
approach, one assumes the surface brightness distributions of both the
galaxy and the PSF to have circular symmetry.  With this assumption,
it can be shown that the 2D seeing convolution is reduced to a one
dimensional integral, with the integrand given by the product of the
surface brightness radial profile of the galaxy model with that of the
PSF surface brightness profile, modulated by a zero-order modified
Bessel function (see BPZ82 for details).  Drawbacks and advantages of
the one and two-dimensional methods have been discussed in many papers
(see \citealt{Kelson:00}, \citealt{LaB:02} and references therein).  To
summarize, the one dimensional approach allows one to significantly
reduce the computation time of galaxy structural parameters.  However,
the uncertainties in the 1D parameters are larger, due to the circular
symmetry approximation as well as to the interpolation of intensity
values which is required to derive the galaxy and PSF one-dimensional
profiles.  On the other hand, the 2D approach is more time consuming,
but allows more accurate estimates of structural parameters by taking
advantage of all the information contained in the galaxy image.  The
one dimensional fitting procedure is included in the 2DPHOT package
for completeness, particularly for cases where galaxy
isophotes are strongly distorted and this distortion changes as a
function of galaxy radius.  In these situations the 2D approach can
provide a poorly constrained fitting model, while useful parameters
can still be obtained by the 1D approach.  In order to apply the BPZ82
method, for each cell of the two-dimensional grid over which the two
dimensional PSF modeling is done (see Sec,~\ref{PSF}), a one
dimensional PSF model is computed.  To this end, the circular surface
brightness profiles of all the sure stars in a given cell are derived
(see Sec.~\ref{ISFIT}) and averaged together after sky subtraction and
flux scaling.  The 1D combined profiles are fit with a sum of Moffat
or Gaussian functions applying a procedure similar to that described
in Sec.~\ref{PSF}.  The one dimensional structural parameters are then
derived by the BPZ82 method, convolving one dimensional S\'{e}rsic models
with the derived 1D PSF models.  The best fitting 1D parameters, i.e.
the central surface brightness, the effective radius and the S\'{e}rsic
index are then derived using $\chi^2$ minimization with the
Levenberg-Marquardt algorithm.  Examples of one dimensional fitting
results are shown in Fig.~\ref{1DPROF} for the same galaxies as in
Fig.~\ref{IS_PLOT}.

\begin{figure}
\scalebox{0.5}{\includegraphics{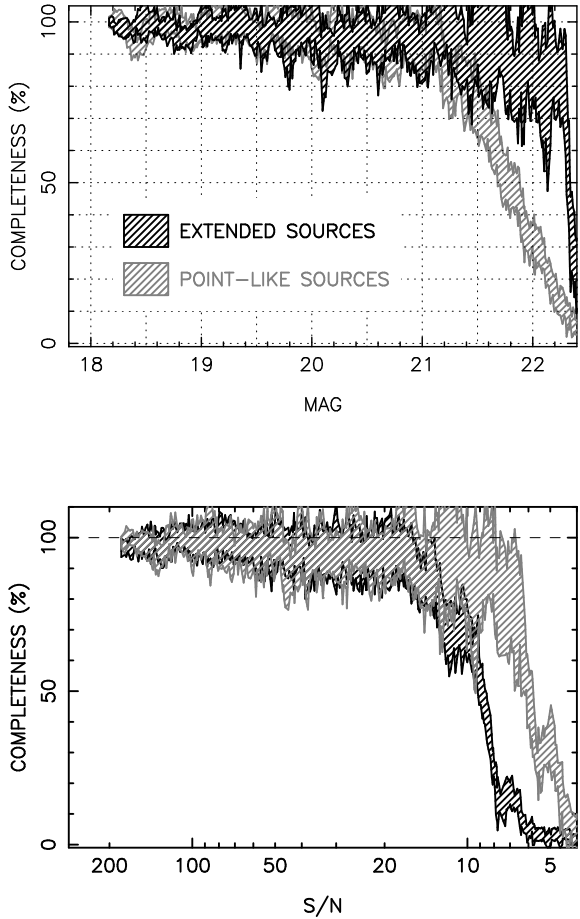}}
\caption[]{\footnotesize Completeness as a function of magnitude
  (upper panel) and S/N ratio (lower panel) for sources in the PACS
  image of the cluster Abell~1081 at $z \sim 0.16$. Hatched regions
  mark $1\sigma$ confidence intervals of the completeness.  Stars and
  galaxies are plotted as grey and black regions, respectively, as
  shown in the upper panel.  The dashed line in the lower plots marks
  the 100$\%$ completeness level.
~\label{COMPL} }
\end{figure}

\section{Growth curves}
\label{GROWTH}
The  aperture magnitude  growth curve  of  each galaxy  is derived  by
direct  integration  of  the  corresponding  one  dimensional  surface
brightness  profile.  The  integration is  performed for  each  of the
concentric ellipses used  to extract the 1D profile.   As described in
Sec.~\ref{SBPROF},  all of  the ellipses  are defined  by  the average
ellipticity and position angle of  the galaxy. In order to correct the
growth  curve   for  seeing   effects,  the  following   procedure  is
adopted. The 2D seeing convolved  S\'{e}rsic model obtained from the final
2D fitting  analysis is used to  extract a growth  curve following the
same  procedure as for  the galaxy  image.  The  growth curve  is also
computed for  the deconvolved  S\'{e}rsic model, which  is defined  by the
output parameters  of the final 2D  fit. The model  is integrated over
concentric ellipses  using an  adaptive 2D integration  algorithm, and
aperture  magnitudes  are extracted  within  the  same apertures  that
define  the galaxy growth  curve.  The  difference between  the growth
curves of the  seeing convolved and the seeing  deconvolved models are
used to  correct the galaxy  aperture magnitudes.  We note  that since
the S\'{e}rsic model  appears in the difference between  the convolved and
the  deconvolved curves,  the  correction is  expected  to be  largely
independent  of  the  choice  of  galaxy  model,  especially  for  the
outermost parts of the galaxy  where seeing corrections are small. The
seeing  corrected growth  curve  is used  to  estimate the  half-light
radius of  the galaxy, and  the corresponding mean  surface brightness
within that radius.  We note  that the seeing corrected growth curve
and surface brightness profile allows the so-called eta function to be
computed,  which is  defined by  the ratio  of the  surface brightness
value at a  given radius to its mean value  within the same radius
(see~\citealt{Sandage:90}).   This  function can  be  used to  compute
Petrosian    metric    radii    and   corresponding    mean    surface
brightnesses. This  feature will be  implemented in  the 2DPHOT
package.

\section{Completeness}
\label{COMPLETENESS}
In order to estimate the completeness of the galaxy catalog, we follow
a procedure similar to that described in Sec.~\ref{SGrules}. 2DPHOT
creates a set of simulated images by adding to the input image a
random spatial distribution of artificial galaxies.  The surface
density of artificial galaxies in each simulated image and the total
number of simulations are chosen using the same criteria outlined in
Sec.~\ref{SGrules}.  Artificial galaxies are created with
seeing-convolved S\'{e}rsic models.  The parameters of each model are
chosen to match the distribution of galaxy structural parameters as a
function of galaxy magnitude obtained from the input image.  The
coordinates of each artificial galaxy are chosen randomly within the
input image, while its total magnitude $m_{\rm g}$ is extracted from a
uniform random distribution spanning the same range as the observed
galaxies\footnote{\footnotesize i.e.  the same magnitude range as objects classified
as extended sources by 2DPHOT}.  2DPHOT randomly selects one of the
fifty objects\footnote{\footnotesize This number is chosen to sample the full range
of galaxy structural parameters at a given galaxy magnitude.  For
several kinds of images, we verified that varying this number from
twenty to eighty does not affect significantly the completeness
estimates.  } in the catalog with magnitudes closest to $m_{\rm g}$
and with corresponding S/N ratio larger than a cutoff value, $S/N_{\rm
min}$.  The INI2DF parameters of this object and the PSF model that
correspond to the extracted center coordinates are then used to create
the artificial galaxy.  The $S/N$ cutoff $S/N_{\rm min}$ is introduced
because at very low S/N ratios the catalog is highly incomplete,
biasing the distribution of galaxy structural parameters toward
objects with a higher detection probability, such as galaxies with
smaller effective radii and/or higher central concentrations (i.e.
higher S\'{e}rsic index).  Since the distribution of galaxy parameters at
magnitudes below the completeness limit is not known, we adopt the
working assumption that this distribution is similar to that of
galaxies which are `close' to the completeness limit of the catalog.
In other words $S/N_{\rm min}$ is chosen as the lowest value of the
$S/N$ ratio for which the catalog is still nearly 100\% complete.
2DPHOT adopts a default value of $S/N_{\rm min}=25$.  However,
processing several images, we found that changing $S/N_{\rm min}$ from
25 to 50 does not significantly change the completeness function.

For each simulated image, a catalog is generated using S-Extractor
with the same settings as for the observed data.  The galaxy
completeness function is then derived by binning the artificial
galaxies in magnitude and measuring the fraction of detected objects
in each bin.  The uncertainties on the completeness function are
estimated by shifting magnitudes of artificial galaxies according to
their corresponding uncertainties and recomputing the fraction of
detected sources in each given bin. The same procedure is applied
to the simulated stars created by 2DPHOT to define
the locus of star candidates (Sec.~\ref{SGrules}).  In this way,
the completeness functions of both extended and point-like sources are
estimated.  Fig.~\ref{COMPL} shows the results of processing one
$r$-band image from the PACS. The figure has been automatically produced
by 2DPHOT. The completeness of the catalog is shown as a
function of both magnitude and $S/N$ ratio. For the latter,
artificial data are binned by $S/N$ ratio and
the fraction of detected sources is measured in $S/N$ bins.  We see
that both the galaxy and the star catalogs are almost $100\%$ complete down to
$S/N \sim 20$ ($r\sim21^m$).

\begin{figure}
\begin{center}
\scalebox{0.43}{\includegraphics{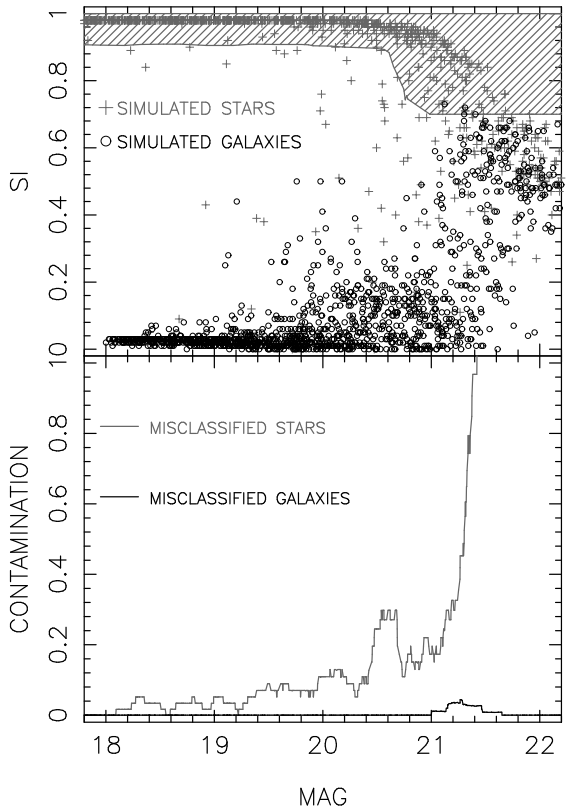}}
\end{center}
\caption[]{\footnotesize The upper panel plots the stellarity index from
  S-Extractor as a  function of the Kron magnitude  of simulated stars
  (grey crosses) and simulated galaxies (black circles), respectively.
  The  hatched  region corresponds  to  the  stellar  locus, which  is
  defined as described in  Sec.~\ref{SGrules}.  The fractions of stars
  and galaxies  that are  erroneously classified on  the basis  of the
  stellar locus  are plotted as  grey and black  curves, respectively.
  The plots  have been  obtained by processing  the PACS image  of the
  cluster Abell~1081, as for Fig.~\ref{COMPL}. ~\label{CONTAM} }
\end{figure}

\section{Contamination}
\label{SG_CONTAM}
\begin{figure*}
\begin{center}
\scalebox{0.7}{\includegraphics{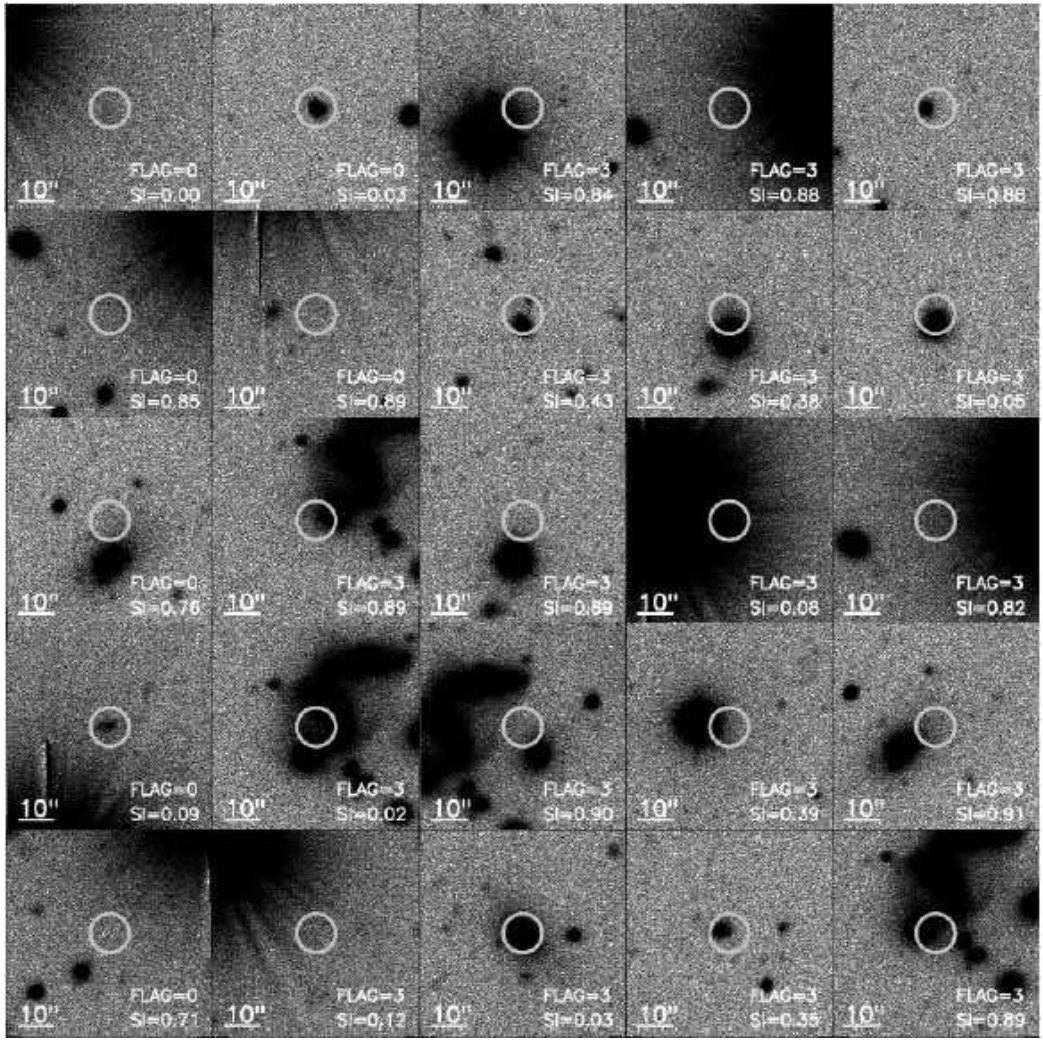}}
\end{center}
\caption[]{\footnotesize  Regions of the same PACS image analyzed in Fig.~\ref{CONTAM} showing
the position of bright simulated stars erroneously classified as galaxies by 2DPHOT (see the text).
Each panel corresponds to a different simulated star. For clarity, we do not show the simulated star images added to the real one, but their positions on each panel are marked by a grey circle with a radius equal to three times the average FWHM value (1.8'') of the PACS
image. The stellarity index and the FLAG parameters estimated by S-Extractor are
reported in the lower-right corner of each panel, while the spatial scale is shown in the lower-left.
~\label{SG_TEST_BRIGHT} }
\end{figure*}

Using simulated stars and galaxies described in Sec.~\ref{SGrules} and
Sec.~\ref{COMPLETENESS},  2DPHOT estimates  the fractions  of galaxies
and stars which  are misclassified as a function  of their magnitudes.
We examine the distribution of the artificial galaxies and stars added
to the  input image  in the $SI$  versus magnitude diagram,  using the
definition  of the  star locus  (Sec.~\ref{SGrules}) to  perform $S/G$
classification.  This  procedure is illustrated  in Fig.~\ref{CONTAM},
where the  results obtained  for one of  the $r$-band PACS  images are
displayed. These  figures are  automatically produced by  2DPHOT.  The
upper panel shows  the star locus as well as  the distribution of both
artificial stars  and artificial galaxies  in the $SI$ -  Mag diagram.
We note that almost all of  the artificial galaxies have $SI \le 0.7$,
which  holds true  for  all  images we  processed  with 2DPHOT.   This
implies that adopting a lower cutoff of $SI=0.7$ for the definition of
the star locus minimizes the fraction of misclassified galaxies at low
$S/N$  ratios, as  noted in  Sec.~\ref{SGrules}.  The  lower  panel of
Fig.~\ref{CONTAM} plots the fraction, $\phi_{\rm s}$, of misclassified
stars, i.e.   the fraction  of artificial stars  that lie  outside the
locus  of  star  candidates,  and  the fraction,  $\phi_{\rm  g}$,  of
misclassified galaxies, i.e.  the fraction of artificial galaxies that
are classified  as stars,  as a function  of their magnitude.   We see
that $\phi_{\rm  g}$ is always smaller  than a few  percent, while the
fraction  of  misclassified  stars  increases  rapidly  at  faint
magnitudes ($MAG>21$).  For  bright  magnitudes, at  $MAG<20$,
where one  would expect  that stars and  galaxies are  always properly
identified, we  find that  the value of $\phi_{\rm  s}$ does
not reach zero,  but is typically $\sim  5 \%$, varying from  $\sim  3\%  $  to $\sim  10\%$  between  $MAG=18$  and
$MAG=20$. In order to understand why there is such a small fraction of
misclassified  stars,  we  considered  the  PACS  frame  whose  2DPHOT
contamination plots are shown  in Fig.~\ref{CONTAM} and selected those
misclassified  stars   for  which  $MAG\le20$,  yielding $25$   out  of  $726$ total stars.
Fig.~\ref{SG_TEST_BRIGHT}  shows the  regions where  each of  these 25
simulated stars are randomly added to the PACS image.  The S-Extractor
stellarity   index  and  FLAG   values  are   also  reported   in  the
plot.  Looking  at the  figure,  we can  clearly  see  that the  small
misclassification  fraction  at the  bright  magnitudes  is caused  by
blending.  In  fact, the figure shows that  bright misclassified stars
can be identified  as follows: (i) they lie just on  top of some other
object in the  field, (ii) strongly blended with  bright galaxies, and
(iii) embedded  within the extended  halo of a bright  saturated star.
We also  find that for $\sim  70 \%$ (18  out of 25) of  the simulated
stars the FLAG  value estimated by S-Extractor is  3, corresponding to
the case  of blended sources  (see ~\citealt{BeA:96}). We  notice that
the blending issue does in principle affect any star/galaxy separation
algorithm, and can be more  or less important depending on how crowded
is  the image  being processed.  On the  other hand,  adding simulated
stars   and  galaxies   to   a   given  image   as   done  by   2DPHOT
(Sec.~\ref{SGrules})   one   can   estimate   the  star   and   galaxy
contamination fractions by  taking into account also misclassification
due to blended sources.

\section{Testing the star/galaxy separation at faint magnitudes}
\label{SG_FAINT}
So far,  we have tested the star/galaxy  separation obtained with
  2DPHOT using images from the  Palomar Abell Cluster Survey. As shown
  in Sec.~\ref{SG_CONTAM}, with PACS  data we achieve reliable star/galaxy
  separation down to  $r \sim 21$. On the  other hand, many scientific
  programs are expected  to reach  significantly deeper limits,
  where the  small size of galaxies and blending issues
  can make  the star/galaxy separation far more  troublesome. In order
  to  discuss how  the star/galaxy  separation in  2DPHOT  performs at
  faint magnitudes, we use two  deep i-band image pointings taken with
  the Large Format  Camera (LFC) at the Palomar  200" telescope.  Each
  LFC pointing  covers a  circular area of  $24'$ in diameter,  with a
  pixel scale of $0.182''/pixel$. Data for the same sky area were also
  taken with the Advanced Camera  for Surveys (ACS) onboard of HST and
  consist of 15  pointings taken with the F814W  filter.  The ACS data
  were drizzled to  a pixel scale of $0.03''/pixel$,  covering a total
  area of $\sim  13 \,arcmin^2$. For more details  on the data quality
  and  main  characteristics  of  the  images,  we  refer  the  reader
  to~\citet{Gal:05}.


\subsection{Comparing the HST and ground-based classification}
\label{LFC_ACS_SG}
\begin{figure}
\begin{center}
\scalebox{0.4}{\includegraphics{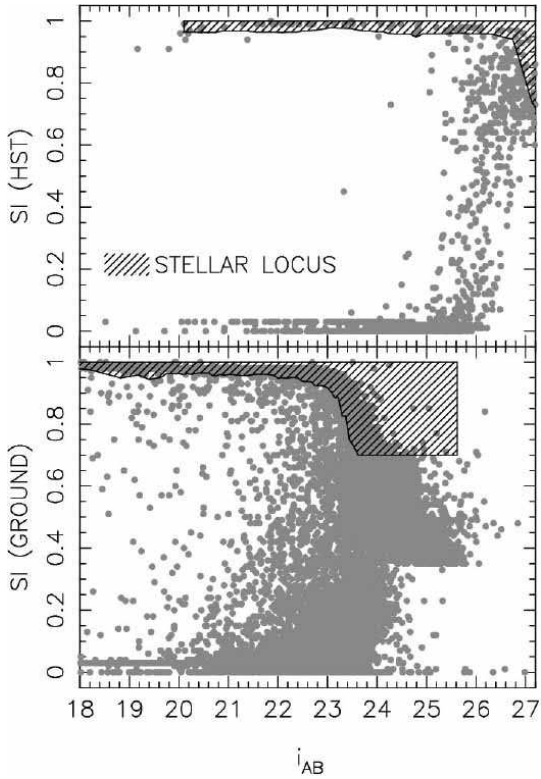}}
\end{center}
\caption[]{\footnotesize  Definition of star locus obtained from 2DPHOT for one
of the HST (upper panel) and one of the LFC (lower panel) images. Each panel plots the
stellarity index versus the $i_{AB}$ magnitude. Grey circles mark all the sources in a given
image, with the corresponding stellar locus being represented by the
hatched region.~\label{STAR_LOCUS_HST_LFC} }
\end{figure}

We  ran 2DPHOT on  each of  the ACS  and LFC  images and  obtained the
corresponding  catalogs of  stars and  galaxies. All  the ACS  and LFC
catalogs were  matched, resulting in a  final list of  3825 sources in
common.  In order to define the star locus, we adopted here the second
option provided by 2DPHOT (see Sec.~\ref{SGrules}), where the locus is
defined  by setting  the minimum  value of  the stellarity  index $SI$
equal to  $\theta - 2.5 \sigma$,  where $\theta$ and  $\sigma$ are the
location  and width  values  of the  $SI$  distribution of  artificial
stars.   As  shown in  Fig.~\ref{STAR_LOCUS_HST_LFC},  where the  star
locus is  plotted for one of  the LFC and  one of the ACS  images, the
above definition  establishes a  narrower stellar region  reducing the
number of small faint  galaxies which can be potentially misclassified
as galaxies.

As shown  in Fig.~\ref{STAR_LOCUS_HST_LFC}, the HST data  go about two
magnitudes  fainter  than the LFC imaging.  Moreover,  HST  allows a  sharp
separation of stars  and galaxies down to $i_{AB}  \sim 25$, while for
LFC the  two classes begin to overlap by $i_{AB} \sim  22$. Assuming
that the HST  data provide the 'true' classification,  we can estimate
the  fraction  of  HST  stars  and galaxies  which  are  not  properly
classified from  LFC and compare these fractions  with those estimated
by   2DPHOT.    Fig.~\ref{STAR_LOCUS_HST_LFC}   compares  the   'true'
misclassified fractions with those predicted by 2DPHOT, as computed by
averaging those  obtained for  the two LFC  fields.  The  figure shows
that  the  2DPHOT  results  are  in good  agreement  with  the  'true'
contamination  estimates. The  fraction of  misclassified  galaxies is
always very  close to zero, reaching $\sim 10\%$ at  $i_{\rm AB} \sim
24$ for both the 'true'  and 2DPHOT estimates. For stars, the fraction
of both 'true' and  2DPHOT misclassified stars increases smoothly with
magnitude, becoming  larger than  $50\%$ at $i_{\rm  AB} \sim  24$. We
notice that the 'true' fraction is slightly larger than that estimated
by 2DPHOT  in the magnitude range  of $i_{\rm AB} \sim  21$ to $i_{\rm
  AB}  \sim 23$. However,  considering the  uncertainty on  the 'true'
fraction  of   misclassified  stars  the  above   difference  is  only
marginally significant.   Hence, we conclude that also  at the fainter
magnitudes sampled  by the LFC  photometry, 2DPHOT is able  to provide
realiable  estimates  of the  contamination  in  the  star and  galaxy
catalogs.

\begin{figure}
\begin{center}
\scalebox{0.45}{\includegraphics{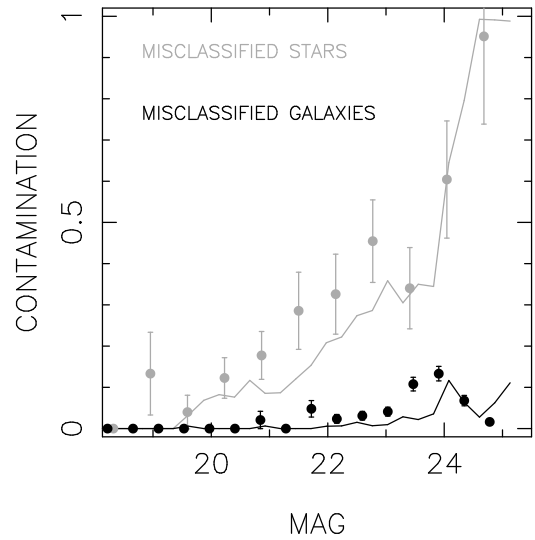}}
\end{center}
\caption[]{\footnotesize 
Comparison of the fractions of misclassified stars and galaxies provided by 2DPHOT
with the 'true' values obtained by comparing the HST and LFC data (see text).
The fractions of stars and galaxies are plotted in grey and black, respectively.
The 'true' fractions are plotted as filled circles, with the error bars marking
$1 \sigma$ standard uncertainties, estimated by accounting for poissonian errors
on counts in each magnitude bin. The fraction of misclassified sources provided by
2DPHOT are shown as continuous curves.~\label{SG_CONTAM_COMB} }
\end{figure}

\section{2DGUI: An interface for 2DPHOT}
\label{INTERF}
The 2DPHOT package requires installation of supporting software
packages\footnote{\footnotesize such as S-Extractor and the cfitsio and pgplot
libraries.} and its performance varies depending on the compilers
used.  After installation, the user has to run the package by
configuring both the input files for S-Extractor as well as some
additional parameters specific to the package itself, which control
the different steps of the image analysis (see Sec.~\ref{2DPHOT}). To
simplify deployment and provide a uniform interface, we have developed
a front-end called 2DGUI. To allow the timely execution of
potentially time-consuming processing jobs and manage parallelization,
we have also included a simple scheduler system.

The 2DGUI package consists of three basic components. First, we
provide an interface where the user can execute several
2DPHOT runs (jobs) through a local 2DPHOT installation. Second, a
small, local (i.e.  server-independent) database provides user access
to the output files of 2DPHOT.  Finally, the simple scheduling system
allows timely execution of several jobs to be performed without server
overloading.  All of these components were either developed or adapted
from well-known, portable, royalty-free software. Since the database is 
included in the 2DGUI package, no additional software
is required. Since the scheduler is based on the {\it cron} utility
and on a bash-- or csh--system shell, the server must use a Unix-like
operating system.  

\begin{figure*}
\begin{center}
\scalebox{0.6}{\includegraphics{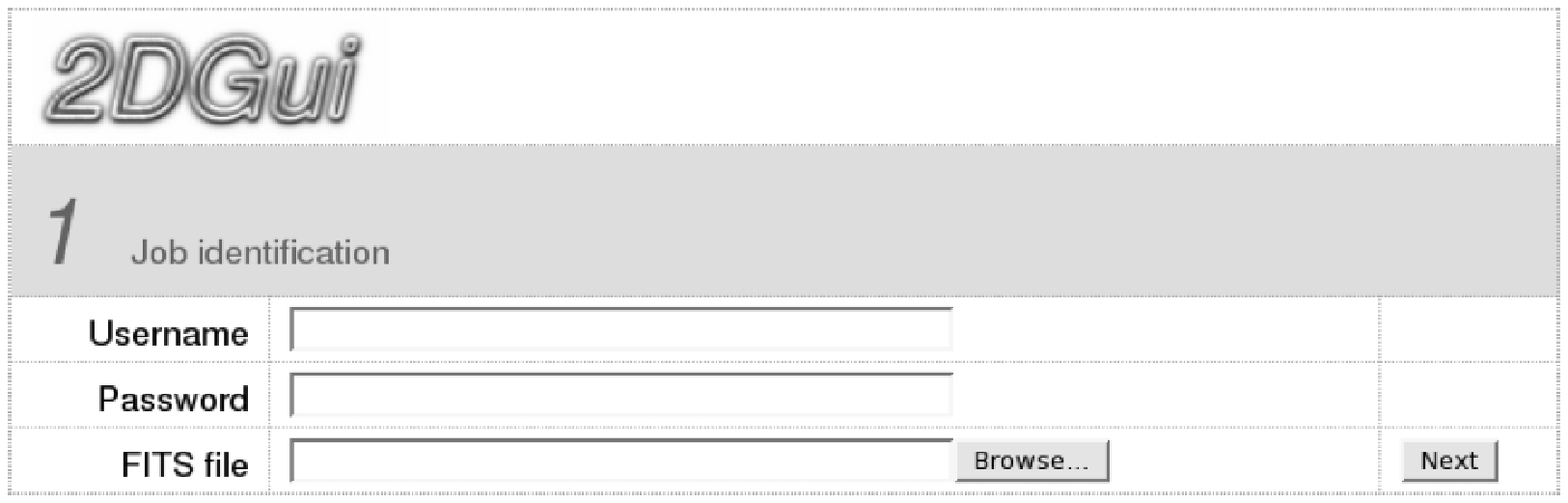}}
\scalebox{0.6}{\includegraphics{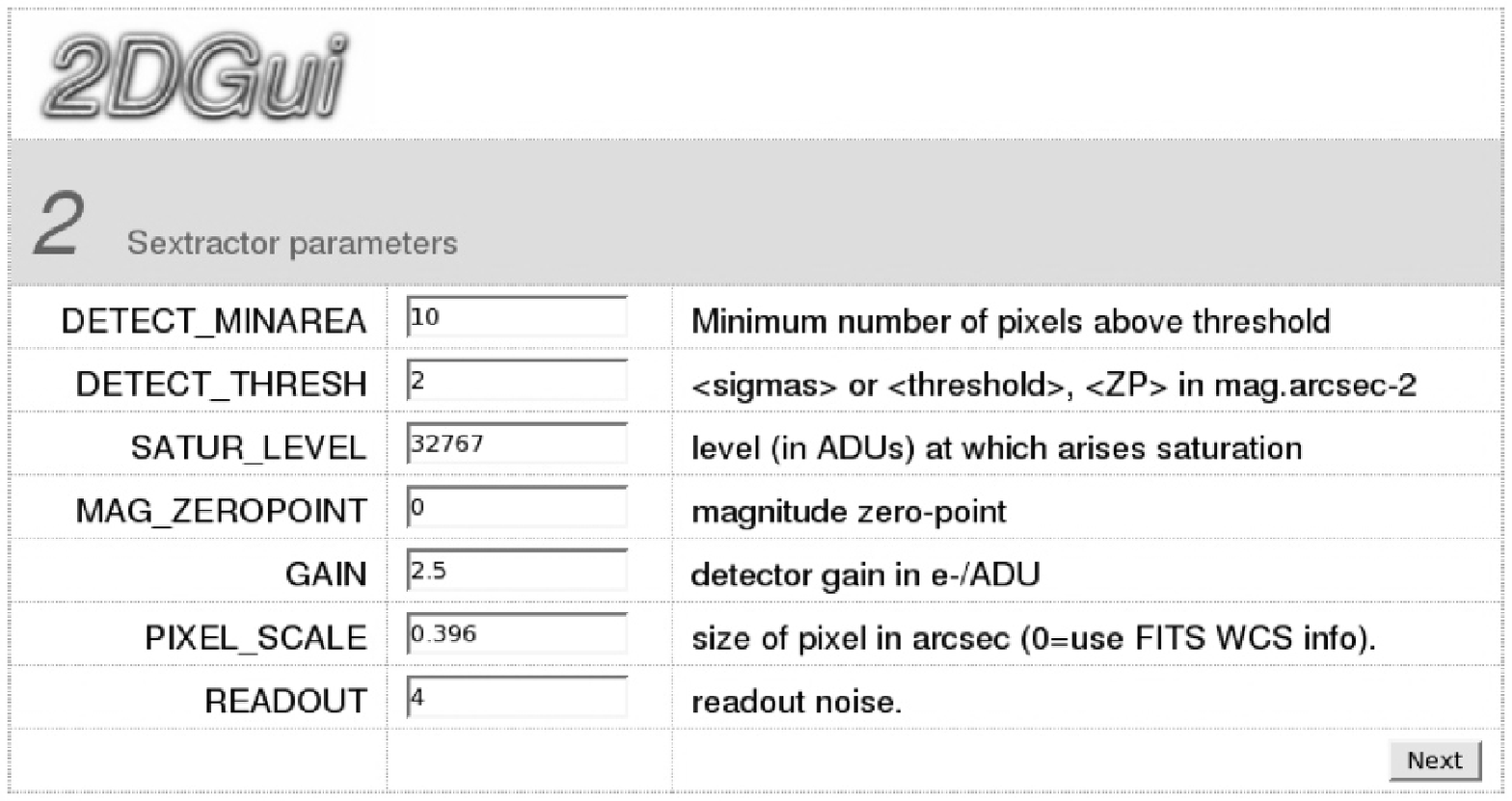}}
\scalebox{0.5}{\includegraphics{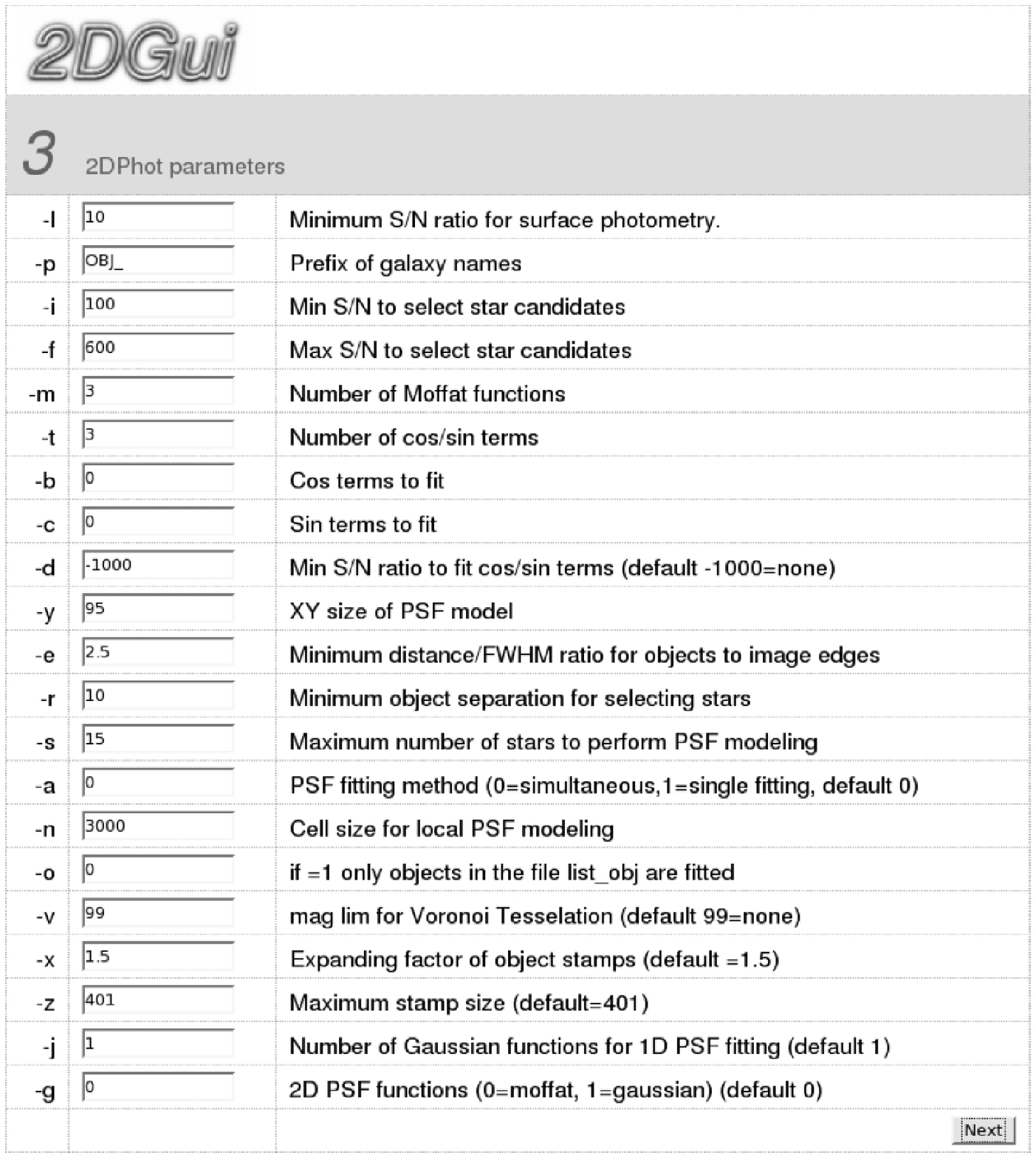}}
\scalebox{0.6}{\includegraphics{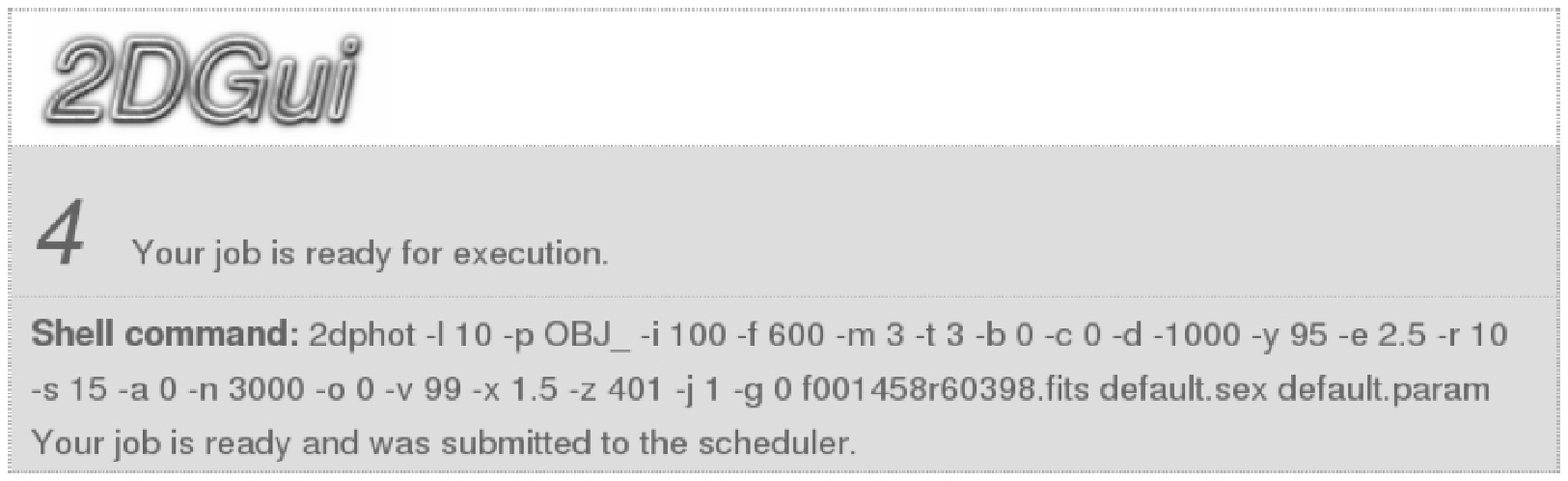}}
\end{center}
\caption[]{\footnotesize  Web forms  used by  the 2DGUI  package.  The
upper form,  panel 1, is  used for the  user login and  image upload.
The middle  forms, panels 2 and 3,  are used to set  the S-Extractor and
2DPHOT parameters. The  bottom form, panel  4, provides
information to the user about the image processing.~\label{panels} }
\end{figure*}

The first step includes user
identification and job creation through the form shown in panel 1 of
Fig.~\ref{panels}.  Currently, only data available on the server can
be processed. After creating a job, the user must configure
parameters for both S-Extractor and 2DPHOT, using the two forms shown
in panels 2 and 3 of Fig.~\ref{panels}, respectively.  Both forms show
the command-line equivalent parameter names, their default values, and
short comments. At this point, the job is created and scheduled, and
information on the job execution is provided in the 2DGUI interface, as shown
in panel 4 of Fig.~\ref{panels}. 

2DGUI then creates a user directory (if necessary), along with a
subdirectory for each job defined by the user identification, the
2DPHOT and S-Extractor parameters, and the filename of the input
image. Four files are stored in this directory: 1) the original FITS
image uploaded by the user; 2) a shell script (runme.sh), that
includes the command-line syntax\footnote{\footnotesize An example of this syntax
is shown in panel 4 of Fig.~\ref{panels}.} for running 2DPHOT; 3 and
4) The S-Extractor configuration and parameter files of S-Extractor,
named {\it default.sex} and {\it default.param}.

During execution, the 2DPHOT main script dumps textual
information on each step of the image analysis in a log file.  2DGUI reads this file and informs the user of
the job processing status by automatically updating the form
shown in panel 4 of Fig.~\ref{panels}.  When processing is finished,
all the files generated by 2DPHOT are listed in the 2DGUI interface, and
the user can select and download 2DPHOT output results.  The
execution of several {\it runme.sh} files on a given server is done by
a scheduling program which runs only a predefined maximum number of
{\it runme.sh} scripts, in such a way as to avoid overloading the server.

Currently, 2DGUI is still under development and its components are
actively being improved.  The major planned improvements include the
following features: 1) the ability for a given user to submit several
jobs, using the same parameters to process several images
simultaneously; 2) use of the local database to store users' preferred
parameters; 3) offline processing alerts (e.g.  sending an e-mail to
the user when a task is completed); and 4) a remote dispatcher, that
allows tasks to be executed remotely, e.g. on a local cluster or a
grid computer. 



\section{Summary}~\label{SUMMARY}  We  have  presented 2DPHOT,  a  new
computational  tool  for astronomical  image  processing, designed  to
analyze  the   output  data  of   wide-field  imaging  surveys   in  a
completely automated  fashion.  The package  includes several tasks,
such as star/galaxy classification, measurement of both integrated and
surface photometry  of galaxies,  PSF modeling, estimation  of catalog
completeness  and  classification accuracy.
2DPHOT  incorporates a variety  of quality  control plots,  which have
historically been  left to a separate  step in image  analysis, and is
complemented  by  a graphical  interface  named 2DGUI.   In
addition, to accommodate the extensive output of 2DPHOT, both in terms
of object  catalogs and quality  control figures, we are  developing a
database architecture which will comply with the standards proposed by
the  IVOA  (International Virtual  Observatory  Alliance).  All  these
components make  2DPHOT a powerful environment to  analyze, handle, and
store the output  data coming from large area  surveys.  Some examples
of surveys  where we plan to  apply this new  analysis environment are
those that  will be carried  out with the VLT-Survey  Telescope (VST).
We emphasize  that 2DPHOT is  conceived as a general  purpose package,
whose possible applications can  span different research topics.  In a
forthcoming publication,  we illustrate one such applications
by  describing a  project  to  measure the  abundance  of clusters  of
galaxies at  high redshifts (up to  $z \sim 1.2$).   Running 2DPHOT on
simulated    images   of    the   VST    KiloDegree    Survey   (KIDS,
see~\citealt{Arn:07}),  we show  that  this cluster  abundance can  be
measured with  a high completeness  level, allowing one to  put strong
constraints on the nature of dark energy in the Universe.

\newpage

\appendix
\section{Input parameters of 2DPHOT}~\label{INPAR}
Table~\ref{INTAB} summarizes the main input parameters of
2DPHOT. These parameters can be either set as input options for the
2DPHOT main script or passed to the package by 2DGUI (see
Secs.~\ref{2DPHOT},~\ref{INTERF}).  In the table, we also include a
short description of each parameter, as well as a reference to the
sections in this paper where the 2DPHOT task influenced by that
parameter is described.

\begin{table*}
\footnotesize
\caption{ Summary of 2DPHOT input parameters. Column 1: Options for
 the 2DPHOT main script. Column 2: Description of the parameter. Column 3: The name used to denote the parameter in the paper text. Column 4: Paper sections 
related to the parameter.
}
\label{INTAB}
\centering
\begin{tabular}{|c|p{12.2cm}|c|c|}
\hline 
    -l &   Minimum S/N ratio required to perform 2D final fitting and surface photometry.  &  & \ref{2DFIT},~\ref{SPHOT}           \\
\hline 
    -x &   Stamp sizes are proportional to the S-Extractor ISOAREA parameter. This parameter provides the proportionality factor. & $EXPND$ & ~\ref{stamps} \\
\hline 
    -z &   Maximum size of the stamp images. This parameter can be used to prevent overly large stamp frames.   &     &  ~\ref{stamps}   \\
\hline 
    -i &   Minimum S/N ratio required to define sure stars. & &\ref{catalog} \\
\hline 
    -f &   Maximum S/N ratio required to define sure stars. & &\ref{catalog} \\
\hline 
    -j &   Number of Moffat/Gaussian functions for 1D PSF fitting. & &\ref{SBPROF} \\
\hline 
    -m &   Number of Moffat/Gaussian functions for 2D PSF fitting. & $NSMAX$ &\ref{PSF}  \\
\hline 
    -g &   Functions used in the 2D PSF fitting (0=Moffat, 1=Gaussian). & & \ref{PSF} \\
\hline 
    -t &   Number of cos/sin terms used for the expansion of star isophotes in the 2D PSF fitting. & & \ref{PSF}\\ 
\hline 
    -d &   Minimum S/N ratio to perform 2D fitting with expansion of the galaxy model into a cos/sin series. & & \ref{2DFIT} \\
\hline 
    -b &   Label providing the cos terms used for the expansion of the galaxy model in  2D  final fitting (e.g. -b 34 makes 2DPHOT calculate the $a_3$ and $a_4$ coefficients) & & ~\ref{2DFIT},~\ref{ISFIT} \\ 
\hline 
    -c &   Label providing the sin terms used for the expansion of the galaxy model in  2D final fitting. & & ~\ref{2DFIT},~\ref{ISFIT} \\ 
\hline 
    -e &   Minimum distance of an object to the image edges, in units of its FWHM. Objects that are closer to the edge this distance are not analyzed. & $REDGE$ & \ref{catalog} \\ 
\hline 
    -s &   Maximum number of sure stars used in a cell to perform PSF modeling.  & $NSIZE$ & \ref{PSF},~\ref{SBPROF}  \\
\hline 
    -a &   Flag that determines the 2D PSF fitting method. When equal to zero, this option forces all sure stars in a given cell to be fitted simultaneously. When equal to one, a single fit to each sure star is performed. & & \ref{PSF} \\
\hline 
    -n &   Size (in pixels) of the grid cells where  PSF modeling is performed.    & & \ref{PSF} \\
\hline 
    -o &   The user can choose to process only some objects in the image by providing a list  of x and y coordinates on the image. This feature is enabled with -o 1. & & \ref{catalog} \\
\hline 
\end{tabular}
\end{table*}

\section{Output quantities measured by 2DPHOT}~\label{OUTPAR}
Table~\ref{OUTTAB} summarizes the output quantities measured by 2DPHOT.
A short  description of  all quantities is  provided, together  with a
reference  to  sections  where  the  corresponding  2DPHOT  tasks  are
described.   The quantities  measured by  running S-Extractor  are not
included for brevity.

\begin{table*}
\footnotesize
\caption{   Summary   of   2DPHOT   output  quantities.    Column   1:
Quantity description. Column 2: Related sections in the text.  }
\label{OUTTAB}
\centering
\begin{tabular}{|p{12cm}|c|}
\hline S-Extractor quantities. & \ref{catalog},~\ref{stamps} \\ 
\hline Stellar locus  quantities: stellar index vs. S/N  ratio for S/G
separation, mean and standard deviation values of the sure star locus. & \ref{catalog},~\ref{stamps},~\ref{SGCLAS}\\ 
\hline PSF  fitting parameters: central intensity,  width, axis ratio,
 position  angle, shape  parameter (in  the case  of PSF  fitting with
 Moffat functions),  and cos/sin terms  of each PSF  fitting function;
 central  coordinates and  local background  value of  each  sure star
 stamp; reduced $\chi^2$ of PSF fitting. These quantities are obtained
 for both the 2D and 1D fitting methods. & \ref{PSF},~\ref{SBPROF}\\
\hline 
 Coarse S\'{e}rsic parameters: center coordinates, central surface brightness, 
effective radius, axis ratio, position angle of the major axis, S\'{e}rsic 
index, total magnitude. & \ref{INI2DF}\\
\hline 
Final  2D  fitting  parameters:  center coordinates,  central  surface 
brightness, effective radius, axis  ratio, position angle of the major
axis, S\'{e}rsic index, total magnitude magnitude, local stamp background value, 
reduced $\chi^2$. & \ref{2DFIT}\\
\hline  
Isophotal  parameters.    For  each  isophote,  the  following
quantities  are  computed:   center  coordinates,  equivalent  radius,
position  angle  of  the  major  axis,  coefficients  of  the  sin/cos
expansion. &  \ref{SPHOT}\\ 
\hline 
1D S\'{e}rsic fitting  parameters: central surface
brightness, effective radius, S\'{e}rsic index, total magnitude,
reduced $\chi^2$. & \ref{SBPROF}\\ 
\hline 
Seeing corrected parameters: aperture magnitudes and surface brightness values 
corresponding to elliptical and circular contours,  half-light radius
and the corresponding mean surface brightness is also computed, petrosian
function. &  \ref{GROWTH}\\ 
\hline
 Completeness function: percentage of recovered simulated stars and galaxies as
a function of the S/N ratio and magnitude. & \ref{COMPLETENESS} \\
\hline 
 Contamination estimates: percentages of misclassified stars and galaxies as
a function of the S/N ratio and magnitude. & \ref{SG_CONTAM} \\
\hline 
\end{tabular}
\end{table*}

\noindent Acknowledgements

We would like to thank Drs. Hugo Capelato, George Djorgovski and Scott
Dodelson for careful reading of  the paper and suggestions that helped
improve  the presentation.   We also  thank Drs.   G.  Busarello, C.P.
Haines,  P.  Merluzzi,  and  M.   Radovich for  helpful  comments  and
suggestions.

\end{document}